\documentclass[sigconf]{acmart}
\renewcommand\footnotetextcopyrightpermission[1]{} % removes footnote with conference information in first column

\AtBeginDocument{%
  \providecommand\BibTeX{{%
    \normalfont B\kern-0.5em{\scshape i\kern-0.25em b}\kern-0.8em\TeX}}}
    
\usepackage{multirow}
\usepackage{amsfonts} 

\usepackage{url}            % simple URL typesetting
\usepackage{booktabs}       % professional-quality tables
\usepackage{amsfonts}       % blackboard math symbols
\usepackage{nicefrac}       % compact symbols for 1/2, etc.
\usepackage{microtype}      % microtypography
\usepackage{lipsum}
\usepackage{bbm}
\usepackage{textcomp}
\usepackage{xcolor}
\usepackage[utf8]{inputenc}
\usepackage[english]{babel}
\usepackage{amsthm}
\usepackage{amsmath}
\usepackage{microtype}
\usepackage{booktabs}
\usepackage{chngpage}
\usepackage{subcaption}
\usepackage{xr}
\usepackage{dsfont}
\usepackage{algorithm,algpseudocode}
\usepackage{hyperref}
\usepackage{wrapfig}
\usepackage{titlesec}

\setcopyright{acmcopyright}
\copyrightyear{2023}
\acmYear{2023}
\acmConference[KDD Marble Workshop 2023]{ACM SIGKDD Conference on Knowledge Discovery and Data Mining Workshop on Multi-Armed Bandits and Reinforcement Learning 2023}{Los Angeles}{CA}
\acmPrice{15.0}
\acmISBN{}

\begin{document}

\title{Scalable Neural Contextual Bandit for Recommender Systems}


% \author{First Author}
% \email{email@domain}
% \affiliation{%
%   \institution{Affiliation}
%   \streetaddress{Address}
%   \city{City}
%   \state{State}
%   \country{Country}
%   \postcode{Postcode}
% }

% \author{Second Author}
% \email{email@domain}
% \affiliation{%
%   \institution{Affiliation}
%   \streetaddress{Address}
%   \city{City}
%   \state{State}
%   \country{Country}
%   \postcode{Postcode}
% }
\author{Zheqing Zhu}
\affiliation{%
  \institution{Meta AI, Stanford Unversity}
  \city{Menlo Park}
  \state{CA}
  \country{USA}
}
\email{billzhu@meta.com}

\author{Benjamin Van Roy}
\affiliation{%
  \institution{Stanford University}
  \city{Stanford}
  \state{CA}
  \country{USA}}
\email{bvr@stanford.edu}
\renewcommand{\shortauthors}{Zhu and Van Roy}

\begin{abstract}
High-quality recommender systems ought to deliver both innovative and relevant content through effective and exploratory interactions with users. Yet, supervised learning-based neural networks, which form the backbone of many existing recommender systems, only leverage recognized user interests, falling short when it comes to efficiently uncovering unknown user preferences. While there has been some progress with neural contextual bandit algorithms towards enabling online exploration through neural networks, their onerous computational demands hinder widespread adoption in real-world recommender systems. In this work, we propose a scalable sample-efficient neural contextual bandit algorithm for recommender systems. To do this, we design an epistemic neural network architecture, Epistemic Neural Recommendation (ENR), that enables Thompson sampling at a large scale.
In two distinct large-scale experiments with real-world tasks, ENR significantly boosts click-through rates and user ratings by at least $9\%$ and $6\%$ respectively compared to state-of-the-art neural contextual bandit algorithms. Furthermore, it achieves equivalent performance with at least $29\%$ fewer user interactions compared to the best-performing baseline algorithm. Remarkably, while accomplishing these improvements, ENR demands orders of magnitude fewer computational resources than neural contextual bandit baseline algorithms.

% require novel and relevant content recommendations through efficient and explorative interactions with users. Supervised learning based neural networks, adopted by most recommender systems, only exploit known interests from users and are not capable of efficiently exploring users' unknown interests. Although state-of-the-art neural contextual bandit algorithms have made progress towards online exploration through deep learning models, many of them require demanding computation and therefore cannot be widely adopted by production recommender systems. In this work, we propose a production-ready sample-efficient neural contextual bandit algorithm for recommender systems.  To do this, we design an epistemic neural network architecture that enables Thompson Sampling at a large scale. We demonstrate through experiments with real-world data that our design improves personalization while maintaining much faster model inference than other neural contextual bandit approaches.
\end{abstract}




\maketitle
\pagestyle{plain} % removes rxunning headers

\section{Introduction}

\begin{figure}[t]
     \centering
     \includegraphics[width=0.4\textwidth]{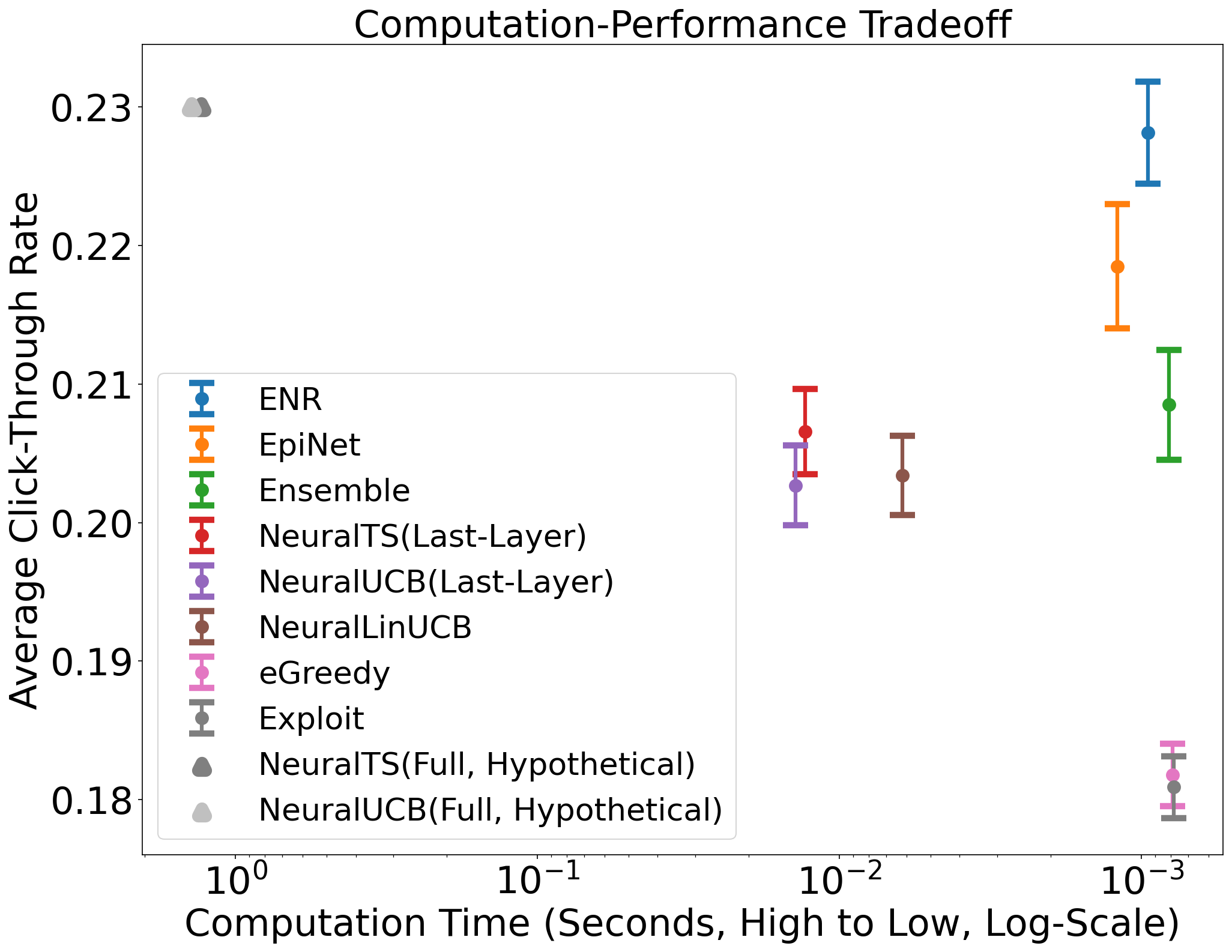}
     \caption{MIND Experiment Computation-Performance Tradeoff. Better Tradeoff towards Top Right. We expect full-scale NeuralUCB and NeuralTS to perform similarly to our methods, but require orders of magnitude more compute. }
     \label{fig:mind_tradeoff}
     \vspace{-0.15in}
 \end{figure}

Recommender systems (RS), paramount in personalizing digital content, critically influence the quality of information accessed via the Internet. Traditionally, these systems have employed supervised learning algorithms, such as Collaborative Filtering \cite{resnick1994grouplens}, which have greatly benefited from advances in deep learning. These algorithms analyze vast quantities of data to discern user preferences; however, they are not designed to strategically probe in order to more quickly learn about user interests. Instead, they learn passively from collected data.

Current research \cite{jiang2019degenerate, salamo2005reducing, gravino2019towards} reveals that deep-learning-driven RS tend to quickly confine their focus to a limited set of suboptimal topics, limiting their scope and hampering their learning capacity. This restrictive personalization strategy confines RS to recommend only those topics with which they have established familiarity, thus failing to discover and learn users' other potential interests. The ability of an RS to identify and learn about user's unexplored interests is a significant determinant of its long-term performance.

The concept of exploration in RS draws heavily from bandit learning \cite{slivkins2019introduction}. In the bandit learning formulation of RS, the system functions as an agent, each user represents a unique context, and each recommendation is an action. Bandit learning algorithms, such as upper confidence bound (UCB) \cite{agrawal1995sample, auer2002finite} and Thompson sampling \cite{thompson1933likelihood}, provide the groundwork for efficient exploration.  Theoretical advances \cite{agrawal2012analysis, dong2018information, dong2019performance, auer2002using, agrawal1995sample, auer2002finite} have offered great insight into these methods, and the efficacy of such methods has been demonstrated through experiments with small-scale environments.  While much of this literature has focused on small-scale environments that do not require that an agent generalizes across contexts and actions, the scale of practical RS calls for such generalization.

Advances in neural bandits offer flexible and scalable approaches to generalization that support more sample-efficient exploration algorithms.  Building on ideas developed for linear UCB and linear Thompson sampling \cite{li2010contextual, chapelle2011empirical}, these approaches offer variations of UCB and Thompson sampling that interoperate with deep learning \cite{li2010contextual, zhou2020neural, xu2021neural,abeille2017linear, zhang2021neural, lu2017ensemble,dwaracherla2019hypermodels,OsbandNeuralTestbed,QinEnsembleSampling}.  While these approaches may be sample-efficient, their computational requirements become onerous at scale.  This has limited their practical impact.

An obstacle to scaling aforementioned approaches has been in the computation required to maintain and apply epistemic uncertainty estimates.  Such estimates allow an agent to {\it know what it does not know}, which is critical for guiding exploration.  The EpiNet \cite{osband2021epistemic} offers a scalable approach to uncertainty estimation, and therefore, a path toward supporting efficient exploration in practical RS.  In particular, epinet-enhanced deep learning combined with Thompson sampling, has the potential to greatly improve RS' exploration capabilities and therefore improve personalization. To this end, we introduce Epistemic Neural Recommendation (ENR), a novel architecture customized for RS. We run a series of experiments using large-scale real-world RS datasets to empirically demonstrate that ENR outperforms state-of-the-art neural contextual bandit algorithms. ENR greatly enhances personalization, achieving an $9\%$ and $6\%$ improvement in click-through rate and user rating, respectively, across two real-world experiments. Furthermore, it attains the performance of the best baseline algorithm with at least $29\%$ fewer user interactions. Importantly, ENR accomplishes these while requiring orders of magnitude fewer computational resources than neural contextual bandit baseline algorithms, making it a considerably more scalable solution. As a spoiler, please see Figure \ref{fig:mind_tradeoff} for a computation-performance tradeoff comparison between our method and baseline methods based on one of our real-world experiments.

% ENR not only captures user preferences more accurately by improving click through rate and user rating by $18\%$ and $6\%$ in two real-world experiments, but also achieves the same performance as the best state-of-the-art algorithm with at least $20\%$ fewer user interactions. Moreover, ENR achieves this performance while requiring orders of magnitude less computational resources.
% Each of these datasets offers full ground truth user feedback for all available user-recommendation pairs and hence no counterfactual evaluation is needed to evaluate algorithm performance. 

The remainder of this paper is organized as follows. In Section 2, we formulate RS as a contextual bandit. In Section 3, we review existing online contextual bandit algorithms, current industry adoptions, and their relative merits. In Section 4, we introduce Epistemic Neural Recommendation. In Section 5, we present empirical results.  In Section 6, we summarize our contributions, benefits they afford, and potential future directions for neural contextual bandits.

\section{Problem Definition} \label{sec:prob}
We conceptualize the design of RS as a contextual bandit problem where an agent interacts with a RS environment by taking an action (recommendation) based on a context (user) observed. More formally, the environment $\mathcal{E}$ is specified by a triple that consists of an observation space $\mathcal{O}$, an action space $\mathcal{A}$, and an observation probability function $P_O$. 

\textbf{Observation space $\mathcal{O}$}: In a contextual bandit problem, the observation generated by the environment offers feedback in response to the previous action taken by the agent as well as a new context the agent needs to execute an action. Formally, $\mathcal{O} = \mathcal{S} \times \mathbb{R}$ and at time step $t$, $O_t = (S_t, R_t)$. $S_t \in \mathcal{S}$ is a new context.  This context could, for example, provide a user's demographic information as well as data from their past interactions.  We will sometimes refer to $S_t$ as a user, since we think of it as providing information that distinguishes users.  $R_t \in \mathbb{R}$ denotes scalar feedback from the user of the previous time step $t-1$ in response to the action $A_{t-1}$.  For example, $R_t$ could be binary-valued, indicating whether the user clicked in response to a recommendation.  We will refer to $R_t$ as the reward received by the agent at time $t$.  

\textbf{Action space $\mathcal{A}$}: At each time $t$, the agent selects an action $A_{t} \in \mathcal{A}$, which identifies a content unit for recommendation to the user $S_t$. The action space $\mathcal{A}$ specifies the range of possibilities afforded by the universe of available content. 
% Given the action and observation definition, we denote History as $H_t = (A_0, O_1, A_1, O_2, A_2, \dots, O_t)$ for simplicity of notation in the rest of the paper. Note that $A_0$ is a dummy action which does not have any consequences from the environment. 

\textbf{Observation probability function $P_O$}: The observation probability function assigns a probability $P_O(O_{t+1}|S_t,A_t)$ to each possible outcome $O_{t+1} = (S_{t+1}, R_{t+1})$ from recommending $A_t$ to user $S_t$.  This probability is a product of two others, specified by functions $P_S$ and $P_R$:
\vspace{-3pt}
\begin{align}
    P_O(O_{t+1} | S_t, A_t) = P_S(S_{t+1})P_{R}(R_{t+1} | S_t, A_t).
\vspace{-5pt}
\end{align}
$P_S$ provides context probabilities, and $P_R$ provides reward probabilities.  At time $t$, the environment samples a new user $S_t \sim P_S$.  The agent then supplies a recommendation $A_t$, and the environment samples $R_{t+1} \sim P_{R}(\cdot| S_t, A_t)$.

The overall objective of the agent is to maximize its average reward $\frac{1}{T}\sum^T_{t=1} R_t$ over $T$ time steps. Note that, though each action $A_t$ does not directly impact any user other than $S_t$, the action influences what information is gathered through the observation $R_{t+1}$.  As such, the action bears an indirect impact on subequent users through what the agent learns, which can improve its ability to make useful recommendations. 
 Optimizing the delayed benefits of learning requires strategic sequential decisions that balance exploration versus exploitation.

We think of $S_t$ and $A_t$ as offering information about users and content units in arbitrary raw formats.  Agents, however, often rely on starting with a more structured representations provided by domain experts.  In this work, we also assume availability of feature extractors that map $S_t$ and $A_t$ to such representations $\phi_{A_t}$ and $\psi_{S_t}$.

Note that our problem formulation encompasses only online learning.  In a real RS, one would pretrain an agent offline on historical data before engaging in online learning.  The methods we develop for online learning can be applied in this workflow post pretraining of an epinet-enhanced model.  However, in order to focus our attention on the problem of exploration, in this paper, we limit our discussions, designs and experiments to the online learning part and leave offline learning for future work.

\section{Related Work and Prerequisites}
Among various bandit algorithms, there are two most commonly adopted online bandit strategies. One explores based on the reward estimates sampled from context-action pairs' posterior distributions, represented by Thompson sampling \cite{thompson1933likelihood}, and the other follows the "optimism in the face of uncertainty" principle, represented by upper confidence bound (UCB) \cite{agrawal1995sample, auer2002finite}. In this section, we offer a high level introduction to Thompson sampling and UCB algorithms as well as their extensions to contextual settings. Their deep neural network versions of the algorithms are considered baselines for our work. We will also review related industry and practical adoptions of these strategies. Note that in this section, we limit our discussion to exploration strategy for immediate reward and do not cover optimization approaches for cumulative returns \cite{zhu2023deep, xu2023optimize, chen2021exploration}.

\subsection{Thompson Sampling, Its Extensions and Related Adoptions} \label{sec:ts}
Thompson sampling is an exploration strategy that samples reward estimates from context-action pairs' posterior distributions and then executes greedily with respect to these samples. This strategy is particularly popular due to its simplicity. Following the definition in Section \ref{sec:prob}, we first consider a vanilla Thompson sampling algorithm \cite{thompson1933likelihood} for multi-armed bandits. For a multi-armed bandit, the context $S_t$ is always the same and the agent needs to choose $A_t$ from $\mathcal{A}$ without any representation learning. A Thompson sampling agent keeps a reward posterior distribution $\hat{P}_{a}$ for each arm $a$ and updates it over time. At time step $t$, the agent chooses
\begin{equation}
    A_t = \arg\max_{a \in \mathcal{A}} \hat{R}_{t+1, a}, \qquad \hat{R}_{t+1, a} \sim \hat{P}_a,
\end{equation}
where $\hat{R}_{t, a}$ is the estimated reward sampled from the agent's posterior belief. The agent then updates its posterior belief $\hat{P}_{a}$ with the latest reward $R_{t+1}$ and selected action $A_t$. The vanilla Thompson sampling approach is a popular production strategy when it comes to small action space recommendations \cite{sanz2019simple, hsieh2015efficient}. 

Extending Thompson sampling to large action space and contextual bandits, an agent can compute parametric estimates of rewards of context-action pairs by sampling linear model parameters from the posterior belief \cite{abeille2017linear, agrawal2013thompson,russo2013eluder,russo2014posterior,russo2018tutorial} instead of sampling a point estimate from the reward posterior distribution. Here, a linear Thompson sampling agent keeps track of a posterior belief $\hat{P}_\theta$ over the parameters $\theta$. The agent chooses action
\begin{equation}
    A_t = \arg\max_{a \in \mathcal{A}} \hat{\theta}_t^\top x_{t, a}, \qquad \hat{\theta}_t \sim \hat{P}_\theta, x_{t, a} = \text{concat}(\psi_{S_t}, \phi_{a}). 
\end{equation}
The agent then updates its posterior belief of $\hat{P}_\theta$ with the latest reward $R_{t+1}$, context $S_t$ and action $A_t$. The posterior update could be performed via Laplace approximation \cite{laplace1986memoir}, and is also presented in industry adoptions of linear Thompson sampling \cite{chapelle2011empirical}. 

The method above could be extended to neural networks as the agent can also keep a posterior belief over neural network parameters. To avoid heavy computation of Laplace approximation, various Bayesian methods have been proposed to achieve posterior updates for distributions over neural network parameters, including deep ensemble \cite{lakshminarayanan2017simple}, ensemble sampling with prior networks \cite{lu2017ensemble}, Bayes by Backprop \cite{blundell2015weight}, Hypermodels \cite{dwaracherla2019hypermodels}, Monte-Carlo Dropouts \cite{gal2016dropout} and EpiNet \cite{osband2021epistemic}. Among the methods above, the two ensemble methods \cite{broden2018ensemble, lu2018efficient, tang2014ensemble} and Monte-Carlo Dropout \cite{guo2020deep} are the most commonly adopted methods for contextual bandits with parameter posterior sampling in a practical setting.

One additional line of work that leverages Thompson sampling for neural networks is neural Thompson sampling \cite{zhang2021neural}. In this approach, instead of sampling from parameters' posterior distribution, neural Thompson sampling directly samples from the posterior estimate of the reward of a context-action pair. A Neural Thompson sampling agent keeps track of a matrix $\Gamma$ initialized with $\Gamma = \lambda \mathbf{I}$, $\mathbf{I} \in \mathbb{R}^{N \times N}$ is an identity matrix, where $\lambda > 0$ is a regularization parameter and $N$ is the parameter size of the neural network. Assuming a Gaussian distributed reward posterior, the variance of the distribution is estimated by
\begin{equation}
    \sigma^2_{t, A} = \lambda g_{\theta}^{\top}(x_{t, A})\Gamma^{-1}g_{\theta}(x_{t, A}),
\end{equation}
where $g_\theta$ is the gradient of the output with respect to $\theta$ for the entire neural network. The agent then chooses action
\begin{equation}
    A_t = \arg\max_{a \in \mathcal{A}}\hat{R}_{t+1, a}, \qquad \hat{R}_{t, a} \sim \mathcal{N}(f_{\theta}(x_{t, a}), \alpha \sigma^2_{t, a}),
\end{equation}
where $\alpha$ is an exploration hyperparameter and $f_{\theta}$ is the neural network. The agent updates $\Gamma \leftarrow \Gamma + g_{\theta}(x_{t, A})g_{\theta}(x_{t, A})^\top / m$ with $m$ being the width of the neural network, assuming all layers with the same width. To the best of the authors' knowledge, the neural Thompson sampling strategy is not well adopted by industry due to its heavy computational complexity. 

\subsection{Upper Confidence Bound (UCB), Its Extensions and Related Adoptions}
Upper confidence bound (UCB) is a general optimism facing uncertainty exploration strategy where the agent tends to choose actions that it has more uncertainties about. The purpose of doing so is to gather information that could help either identify the best action or eliminate bad actions. Similar to the subsection above, we first consider a vanilla UCB algorithm \cite{agrawal1995sample, auer2002finite} for multi-armed bandits. A vanilla UCB agent selects an action $A_t$ by 
\begin{equation}
    A_t = \arg\max_{a \in \mathcal{A}} \left(\frac{\sum_{\tau=1}^{t} R_{\tau+1} \mathbbm{1}(A_\tau = a)}{n_{t, a}} + \alpha\sqrt{\frac{\ln(t)}{n_{t, a}}}\right),
\end{equation}
where $n_{t, A} = \sum_{\tau = 1}^t \mathbbm{1}(A_\tau = A)$ and $\alpha \in \mathbb{R_+}$ is a hyperparameter. There have been multiple lines of work analyzing vanilla UCB's impact on production recommender systems in terms of degenerating feedback loops \cite{jiang2019degenerate, Guo2023evaluate}, modeling attritions \cite{ben2022modeling} and its extension to Collaborative Filtering \cite{nakamura2015ucb}.

The immediate extension to UCB to linear bandit problems is LinUCB \cite{li2010contextual, chu2011contextual}. A LinUCB agent initializes two variables for any new action $A$ that it has never seen before, $\Gamma_A = \mathbf{I}_d$ ($d$-dimensional identity matrix) and $b_A = \mathbf{0}_{d\times 1}$ ($d$-dimensional zero vector), where $d$ is the linear parameter size. At time step t, given a new context representation $\psi_{S_t}$ and each action $\phi_{A}$, the agent concatenates the two vectors to get a context-action pair representation $x_{t, A} = \text{concat}(\psi_{S_t}, \phi_A)$. The agent selects action
\begin{equation}
    A_t = \arg\max_{a \in \mathcal{A}}\left((\Gamma^{-1}_a b_a)^\top x_{t, a} + \alpha \sqrt{x_{t, a}^\top\Gamma_a^{-1}x_{t, a}}\right).
\end{equation}
After receiving $R_{t+1}$, the agent then updates $\Gamma_{A_t}$ and $b_{A_t}$ with $\Gamma_{A_t} \leftarrow \Gamma_{A_t} + x_{t, A_t}x_{t, A_t}^\top$ and $b_{A_t} \leftarrow b_{A_t} + R_{t+1} x_{t, A_t}$. LinUCB is a popular algorithm for adoption in many industry usecases and particularly in the space of recommender systems \cite{li2010contextual, zhao2013interactive, qin2014contextual, wang2017biucb}. 

With the rise of neural networks and deep learning, so comes the need for optimism based exploration neural methods. NeuralUCB \cite{zhou2020neural} and NeuralLinUCB \cite{xu2021neural} are two representing examples of such algorithms. We first introduce NeuralLinUCB as it's a natural extension of LinUCB. Given a neural network representation, NeuralLinUCB no longer initiates a matrix $\Gamma_A$ for each action $A$ and instead keeps track of a single matrix $\Gamma$ initialized with $\lambda \mathbf{I}$ where $\lambda > 0$ is a regularization factor and $\mathbf{I} \in \mathbb{R}^{N \times N}$ is an identity matrix. Suppose the neural network's last layer representation is $\sigma_{\theta}(x_{t, A}) \in \mathbb{R}^N$ and full model output $f_\theta(x_{t, A}) \in \mathbb{R}$, the agent takes action
\begin{equation}
    A_t = \arg\max_{a \in \mathcal{A}}\left(f_{\theta}(x_{t, a}) + \alpha \sqrt{\sigma_{\theta}(x_{t, a})^\top\Gamma^{-1}\sigma_{\theta}(x_{t, a})}\right),
\end{equation}
with the same update for $\Gamma$ and $b$ except that now the update happens with the representation $\sigma_\theta(x_{t,A})$ instead of $x_{t,A}$. NeuralUCB follows the same principle as above, but replacing $\sigma_{\theta}(x_{t,A})$ with the gradients of all parameters within the neural network. 

Although empirically the above neural methods, Neural Thompson sampling, Neural UCB and Neural LinUCB, have achieved good performance on synthetic datasets, their downside is that they all require inverting a square matrix with its dimension equal to either the entire neural network's parameter size or the size of the last layer representation. Both in many cases are intractable, especially for real-world environments and complex neural networks. 

\section{Epistemic Neural Recommendation}
In this section, we introduce a novel architecture for scalable neural contextual bandit problems, drawing inspiration from Thompson sampling and recent developments in epistemic neural networks (ENN) and EpiNet \cite{osband2021epistemic}. As discussed in Section \ref{sec:ts}, the objective of a deep learning-based Thompson sampling strategy is to accurately estimate the uncertainty of a prediction, pertaining to a context-action pair, while keeping computational costs to a minimum. A significant drawback of many neural methods for Thompson sampling and UCB is their computationally expensive uncertainty estimation process, which impedes their integration into real-world environments. The primary aim of our proposed architecture and design is to address this very issue, while enhancing model performance.

\subsection{Informative Neural Representation}
A key aspect shared across all neural exploration methods is the generation of effective representations for context-action pairs. Within the contextual bandit framework, there are three main representation factors to consider: action representation, context representation, and the interaction between context and action.

We assume a general and unnormalized feature representation from the environment for both contexts and actions. To distill both context and action feature representation vectors, we initially feed each vector into a linear layer followed by a ReLU activation function. This is subsequently followed by a Layer Normalization (LayerNorm) layer \cite{ba2016layer}. Given that raw features can contain disproportionately large values that are challenging to normalize - particularly in an online bandit setting where data continually stream in - it is crucial to employ layer normalization. This technique ensures smoother gradients, promoting stability and generalizability \cite{xu2019understanding}. More formally, 
\begin{equation}
\begin{split}
    h_{\beta_{\text{context}}}(\psi_{S_t}) &= \text{LayerNorm}(\text{ReLU}(\beta_{\text{context}}^\top\psi_{S_t})),\\
    h_{\beta_{\text{action}}}(\phi_A) &= \text{LayerNorm}(\text{ReLU}(\beta_{\text{action}}^\top\phi_A)),
\end{split}
\end{equation}
where $\theta_{\text{context}} \in \mathbb{R}^{d_{S} x d_{E}}$ and $\theta_{\text{action}} \in \mathbb{R}^{d_{A} x d_{E}}$ are the parameters for context and action summarization. Note that after summarization through these two layers, the outputs are of the same shape.

Having summarized and normalized representations for both action and context, we now model the interaction between these two entities. Collaborative Filtering \cite{resnick1994grouplens, goldberg1992using, herlocker2000explaining} and Matrix Factorization \cite{koren2009matrix} provide intuitive models for this interaction. Inspired by these methods, we use element-wise multiplication to represent interactions,
\begin{equation}
    I(\psi_{S_t}, \phi_A) = h_{\beta_{\text{context}}}(\psi_{S_t}) \odot h_{\beta_{\text{action}}}(\phi_A).
\end{equation}

Utilizing the above three representations, we concatenate the three vectors to derive

\begin{equation} \label{eq:x}
x_{t, A} = \text{concat}\left(h_{\beta_{\text{context}}}(\psi_{S_t}), h_{\beta_{\text{action}}}(\phi_A), I(\psi_{S_t}, \phi_A)\right).
\end{equation}

Notably, this representation diverges from Neural Collaborative Filtering (NCF) \cite{he2017neural} in two primary ways. Firstly, NCF can only manage id-listed features and it aggregates the concatenation of action and context through multi-layer perceptrons. In contrast, our approach can handle feature vectors without making any preliminary assumptions about the input feature shape or values. Secondly, we use the representations obtained through summarization for uncertainty estimation, as discussed in the following subsection. Moreover, we demonstrate in later sections through empirical studies that the direct application of NCF with uncertainty estimation yields inferior performance compared to our method.

\subsection{Epistemic Neural Recommendation (ENR)}
Using the representation derived above, we can now design a neural network for both point estimate and uncertainty estimation. For efficient epistemic uncertainty estimation using neural networks, we employ EpiNet \cite{osband2021epistemic}, an auxiliary architecture for a general neural network. This architecture leverages the final layer representation of a neural network, along with an epistemic index $z$, to generate a sample from the posterior. EpiNet presents a cost-effective method for constructing ENNs, which has been demonstrated to excel in making joint predictions for classification and supervised learning tasks, while achieving strong performance in synthesized neural logistic bandits with minimal additional computational costs. However, a drawback of only utilizing the last layer representation, particularly in a contextual bandit setting, is the diminished representation power. The interactions between context and action are summarized through multi-layers perceptrons for marginal predictions, and considerable information is lost in the process. Motivated by this, we introduce Epistemic Neural Recommendation (ENR) that generates improved posterior samples via more informative representations.

The remaining architecture of ENR is comprised of three parts: a function $f_{\theta_x}: \mathbb{R}^{3d_E} \mapsto \mathbb{R}$ for marginal prediction, and two functions $g_{\sigma}$ and $g_{\sigma^p}$, both mapping from $\mathbb{R}^{3d_E}$ to $\mathbb{R}^{d_z}$, for uncertainty estimation. Upon extraction of $x_{t, A}$, we employ $f_{\theta_x}$ to make the marginal prediction via $f_{\theta_x}(x_{t,A})$.

In addition to $f_{\theta_x}$, $g_{\sigma}$ and $g_{\sigma^p}$ are independently initialized through Glorot Initialization \cite{glorot2010understanding, bartlett2017spectrally}, and $g_{\sigma^p}$ remains constant throughout its lifetime, thereby providing robust regularization. To initialize this segment of the network, ENR samples an epistemic index $z$ from its prior $P_z$, which could either be a discrete one-hot vector or a standard multivariate Gaussian sample. The sampled $z$ is then concatenated with $x_{t, A}$ and processed through both $g_{\sigma}$ and $g_{\sigma^p}$. Consequently, the uncertainty estimation is $(g_{\sigma}(x_{t, A}, z) + g_{\sigma^p}(x_{t, A}, z))^\top z$. Hence the final output of ENR is
\begin{equation} \label{eq:forward}
    f_\theta(\psi_S, \phi_A, z) = f_{\theta_x}(x_{t, A}) + (g_{\sigma}(sg[x_{t, A}], z) + g_{\sigma^p}(sg[x_{t, A}], z))^\top z,
\end{equation}
where $x_{t, A}$ is defined in Equation \ref{eq:x}, $\theta = (\theta_x, \sigma, \sigma^p, \beta_{\text{context}}, \beta_{\text{action}})$, and $sg[\cdot]$ stops gradient flow of the parameter within the bracket. Note that, if the reward function is a binary variable, then a sigmoid would be added to the final output of $f_\theta(\psi_S, \phi_A, z)$ as neural logistic regression is a common setup in RS. 

With the neural network above, an ENR Thompson sampling agent executes action 
\begin{equation}
    A_t = \arg\max_{a \in \mathcal{A}}f_\theta(\psi_{S_t}, \phi_a, z), z \sim P_z.
\end{equation}
After receiving reward $R_{t+1}$, then the agent can update the neural network with 
\begin{equation}
    \theta \leftarrow \theta - \alpha\nabla_{\theta}\sum_{z\in \mathcal{Z}}\mathcal{L}(R_{t+1}, f_\theta(\psi_{S_t}, \phi_{A_t}, z))
\end{equation}
through stochastic gradient descent, where $\mathcal{Z}$ is a set of epistemic indices sampled from $P_z$ and $\alpha$ is the learning rate. In practice, we perform batch updates through sampling from a replay buffer and uses other more advanced optimization strategies such as ADAM \cite{kingma2015adam} for parameter updates. The loss function $\mathcal{L}$ here could be cross-entropy loss for neural logistic contextual bandits and mean squared error loss for contextual bandits with non-binary reward. 

\begin{figure}[t]
     \centering
     \includegraphics[width=0.53\textwidth]{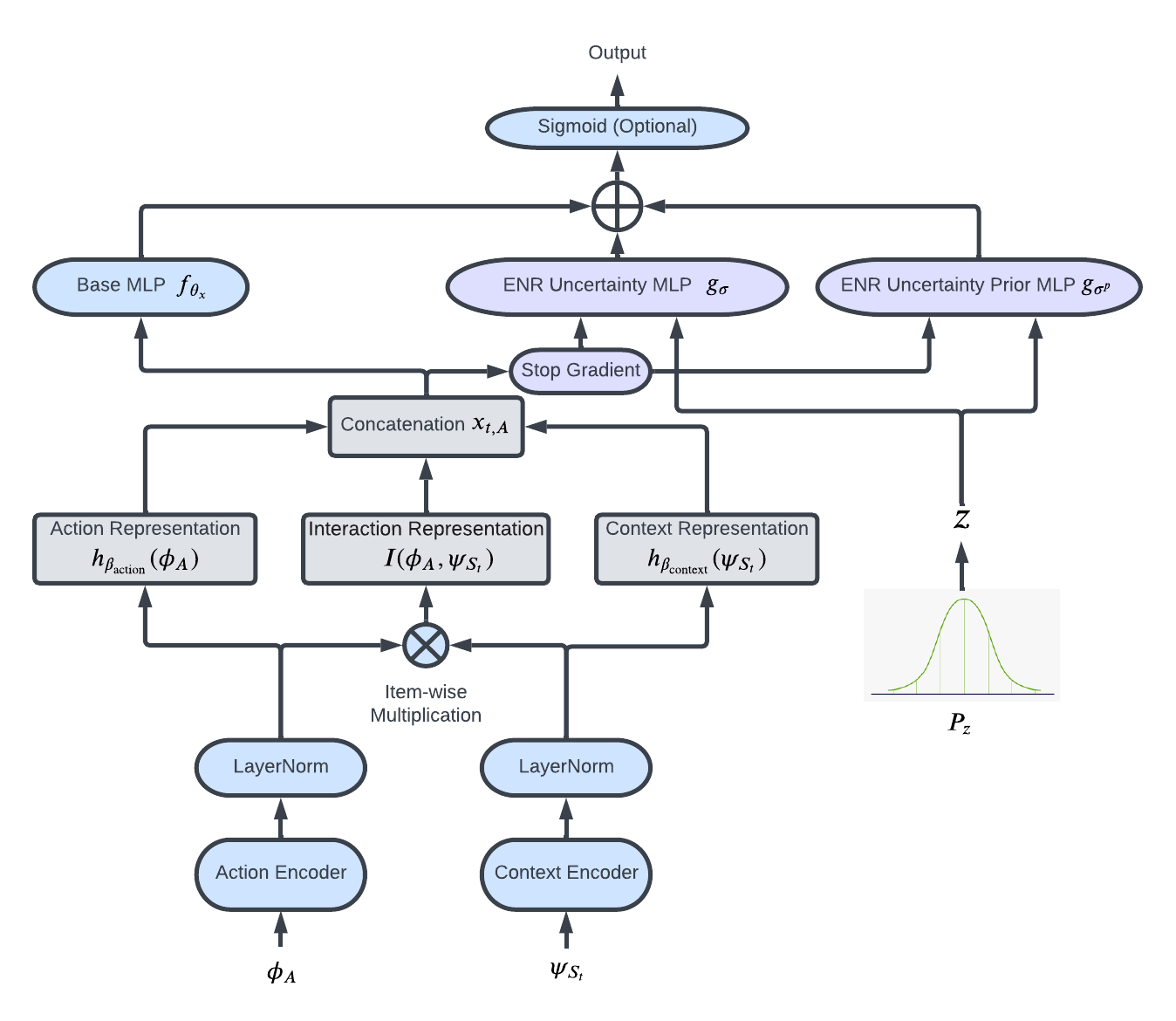}
     \caption{Epistemic Neural Recommendation Architecture}
     \label{fig:enr}
     \vspace{-0.1in}
\end{figure}

For the full illustration of the architecture, please refer to Figure \ref{fig:enr}. For a detailed algorithm illustration, please refer to Algorithm 1. We would like to also note that this algorithm is a full online bandit algorithm without any prior knowledge or pre-training. This algorithm can also be extended to perform offline pretraining before online interactions to reduce the amount of exploration required. We leave this for future work. 

\begin{algorithm}[t]
\small
	\caption{Epistemic Neural Recommendation (ENR) Thompson Sampling}\label{alg:enr}
	\begin{algorithmic}[1]
	    \State Initialize $\theta, P_z$.
	    \State Initialize a replay buffer $\mathcal{D}$
		\For{$t = 1, 2, \dots$} \Comment{Progressing over time steps}
    		\State Makes observation $O_{t}$, obtains $\psi_{S_t}$ and computes $R_{t}$.
    		\State Store $(\psi_{S_{t-1}}, \phi_{A_{t-1}}, R_t)$ in $\mathcal{D}$.
    		\State Sample $z \sim P_z$ and compute $f_\theta(\psi_{S_t}, \phi_{A}, z) \forall A$ according to Equation \ref{eq:forward}.
    		\State $A_t = \arg\max_{a \in \mathcal{A}}f_{\theta}(\psi_{S_t}, \phi_a, z)$.
    		\For{$k = 1, 2, \dots, K$} \Comment{Optimization}
    	        \State Sample a data batch $B \sim \mathcal{D}$.
    	        \State Sample a set of epistemic indices $Z \sim P_z$
    	        \State $\theta \leftarrow \theta - \alpha\nabla_{\theta}\sum_{i \in B}\sum_{z\in Z}\mathcal{L}(R_{i}, f_\theta(\phi_{S_{i}}, \phi_{A_{i}}, z))$.
    	    \EndFor
		\EndFor
    \end{algorithmic}
    \vspace{-0.03in}
\end{algorithm}

\section{Experiments}
\begin{table}[t!]
\small
    \centering
  \caption{Experiment Dataset Statistics}
  \vspace{-0.1in}
  \label{tab:stats}
  \begin{tabular}{cccccccccc}
    \toprule
    Dataset & \# Users & \# Items & \# Interactions\\
    \midrule
    MIND \cite{wu2020mind} & 1,000,000 & 160,000 & 15,000,000\\
    KuaiRec \cite{gao2022kuairec} & 1,411 & 3,327 & 4,676,570\\
  \bottomrule
\end{tabular}
% \vspace{-0.3in}
\end{table}

\begin{table*}[h]
\small
  \caption{MIND Dataset Illustration}
  \vspace{-0.1in}
  \label{tab:mind_example}
  \begin{tabular}{ccccc}
    \toprule
    Impression ID & User ID & Time & User Interest History & News with Labels\\
    \midrule 
    91 & U397059 & 11/15/2019 10:22:32 AM & N106403 N71977 N97080 N102132 & N129416-0 N26703-1 N120089-1 N53018-0 N89764-0\\
  \bottomrule
\end{tabular}
\vspace{-0.15in}
\end{table*}

\begin{table}[t!]
\small
  \caption{Parameter Sizes across Different Architectures}
  \vspace{-0.1in}
  \label{tab:param}
  \begin{tabular}{ccccc}
    \toprule
    Algorithm & MIND & KuaiRec\\
    \midrule 
    MLP & $0.76 \times 10^6$ & $1.21 \times 10^7$\\
    Wide \& Deep & $/$ & $2.33 \times 10^7$\\
    NCF & $1.51 \times 10^6$ & $2.31 \times 10^7$\\
    NRMS & $1.49 \times 10^6$ & $/$\\ 
    NAML & $1.49 \times 10^6$ & $/$\\
    DeepFM & $/$ & $2.30 \times 10^7$ \\
    EpiNet + MLP & $0.77 \times 10^6$ & $1.22 \times 10^7$\\
    EpiNet + Wide \& Deep & $/$ & $2.34 \times 10^7$\\
    EpiNet + NCF & $1.52 \times 10^6$ & $2.32 \times 10^7$\\
    EpiNet + NRMS & $1.50 \times 10^6$ & $/$ \\
    EpiNet + NAML & $1.50 \times 10^6$ & $/$\\
    EpiNet + DeepFM & $/$ & $2.31 \times 10^7$ \\
    ENR & $1.52 \times 10^6$ & $2.32 \times 10^7$ \\
  \bottomrule
\end{tabular}
\vspace{-0.1in}
\end{table}

\begin{table}[t!]
\small
  \caption{Average Computation Time for MIND Experiments (Seconds). Both full Neural TS and Neural UCB agents require inverting a covariance matrix of approximate size $10^{6} \times 10^{6}$, which leads to $\infty$ inference time and intractable memory. Even when only using the last layer gradients, Neural TS and Neural UCB still takes about 10 times of ENR agent's inference time.}
  \vspace{-0.1in}
  \label{tab:mind_comp}
  \begin{tabular}{ccccc}
    \toprule
    Algorithm & Inference / Action & Training\\
    \midrule 
    Neural UCB (w/ any arch. below) & $\infty$ & $\infty$\\
    Neural TS (w/ any arch. below) & $\infty$ & $\infty$\\
    \midrule
    $\epsilon$-Greedy + MLP  & $0.0006 \pm 0.0001$ & $0.11 \pm 0.01$\\
    Neural UCB (Last Layer) + MLP & $0.0098 \pm 0.001$ & $0.13 \pm 0.01$\\
    Neural LinUCB + MLP & $0.0045 \pm 0.001$ & $0.13 \pm 0.01$\\
    Neural TS (Last Layer) + MLP & $0.0099 \pm 0.001$ & $0.14 \pm 0.01$\\
    Ensemble + MLP & $0.0006 \pm 0.0001$ & $0.72 \pm 0.03$\\

    \midrule

    $\epsilon$-Greedy + NCF & $0.00079 \pm 0.0001$ & $0.13 \pm 0.01$\\
    Neural UCB (Last Layer) + NCF & $0.014 \pm 0.002$ & $0.16 \pm 0.01$\\
    Neural LinUCB + NCF & $0.0062 \pm 0.001$ & $0.15 \pm 0.01$\\
    Neural TS (Last Layer) + NCF & $0.013 \pm 0.002$ & $0.16 \pm 0.01$\\
    Ensemble + NCF & $0.00081 \pm 0.0001$ & $0.93 \pm 0.03$\\

    \midrule
    
    $\epsilon$-Greedy + NAML & $0.00082 \pm 0.0001$ & $0.13 \pm 0.01$\\
    Neural UCB (Last Layer) + NAML &$0.014 \pm 0.002$ & $0.15 \pm 0.01$\\
    Neural LinUCB + NAML &$0.063 \pm 0.001$ & $0.15 \pm 0.01$\\
    Neural TS (Last Layer) + NAML & $0.014 \pm 0.002$ & $0.15 \pm 0.01$\\
    Ensemble + NAML & $0.00079 \pm 0.0001$ & $0.95 \pm 0.03$\\

    \midrule

  $\epsilon$-Greedy + NRMS & $0.00083 \pm 0.0001$ & $0.14 \pm 0.01$\\
    Neural UCB (Last Layer) + NRMS & $0.013 \pm 0.002$ & $0.16 \pm 0.01$\\
    Neural LinUCB + NRMS & $0.006 \pm 0.001$ & $0.15 \pm 0.01$\\
    Neural TS (Last Layer) + NRMS & $0.014 \pm 0.002$ & $0.15 \pm 0.01$\\
    Ensemble + NRMS & $0.00078 \pm 0.0001$ & $0.91 \pm 0.03$\\

    \midrule\midrule
    EpiNet + MLP & $0.0007 \pm 0.0001$ & $0.12 \pm 0.01$\\
    EpiNet + NCF & $0.0012 \pm 0.0001$ & $0.16 \pm 0.01$\\
    EpiNet + NAML & $0.0013 \pm 0.0001$ & $0.17 \pm 0.01$\\
    EpiNet + NRMS & $0.0012 \pm 0.0001$ & $0.17 \pm 0.01$\\
    \midrule
    ENR & $0.00095 \pm 0.0001$ & $0.18 \pm 0.01$\\
  \bottomrule
\end{tabular}
\vspace{-0.2in}
\end{table}

\begin{table}[t!]
\small
  \caption{Average Computation Time for KuaiRec Experiments (Seconds). Both full Neural TS and Neural UCB agents require inverting a covariance matrix of approximate size $10^{7} \times 10^{7}$, which leads to $\infty$ inference time and intractable memory. Note that NeuralUCB and NeuralTS agents that only compute the last layer of gradients still takes 150 times of ENR's inference time due to available action size of 3327 for every time step, hence intractable in this experiment.}
  \vspace{-0.1in}
  \label{tab:kuai_comp}
  \begin{tabular}{ccccc}
    \toprule
    Algorithm & Inference / Action & Training\\
    \midrule
    Neural UCB (w/ any arch. below) & $\infty$ & $\infty$\\
    Neural TS (w/ any arch. below) & $\infty$ & $\infty$\\
    \midrule
    $\epsilon$-Greedy + MLP & $0.012 \pm 0.001$ & $0.24 \pm 0.01$\\
    Neural UCB (Last Layer) + MLP & $1.60 \pm 0.13$ & $0.25 \pm 0.01$\\
    Neural LinUCB + MLP & $0.036 \pm 0.002$ & $0.25 \pm 0.01$\\
    Neural TS (Last Layer) + MLP & $1.59 \pm 0.13$ & $0.25 \pm 0.01$\\
    Ensemble + MLP & $0.018 \pm 0.001$ & $1.2 \pm 0.03$\\
    \midrule
    $\epsilon$-Greedy + NCF & $0.013 \pm 0.001$ & $0.31 \pm 0.02$\\
    Neural UCB (Last Layer) + NCF & $1.71 \pm 0.13$ & $0.35 \pm 0.02$\\
    Neural LinUCB + NCF & $0.042 \pm 0.002$ & $0.36 \pm 0.02$\\
    Neural TS (Last Layer) + NCF & $1.72 \pm 0.13$ & $0.36 \pm 0.02$\\
    Ensemble + NCF & $0.020 \pm 0.001$ & $1.5 \pm 0.03$\\
    \midrule
    $\epsilon$-Greedy + Wide \& Deep & $0.013 \pm 0.001$ & $0.32 \pm 0.02$\\
    Neural UCB (Last Layer) + Wide \& Deep & $1.68 \pm 0.13$ & $0.34 \pm 0.02$\\
    Neural LinUCB + Wide \& Deep & $0.039 \pm 0.002$ & $0.35 \pm 0.02$\\
    Neural TS (Last Layer) + Wide \& Deep & $1.69 \pm 0.13$ & $0.34 \pm 0.02$\\
    Ensemble + Wide \& Deep & $0.020 \pm 0.001$ & $1.6 \pm 0.03$\\
    
    \midrule
    $\epsilon$-Greedy + DeepFM & $0.014 \pm 0.001$ & $0.32 \pm 0.02$\\
    Neural UCB (Last Layer) + DeepFM & $1.72 \pm 0.13$ & $0.36 \pm 0.02$\\
    Neural LinUCB + DeepFM & $0.041 \pm 0.002$ & $0.36 \pm 0.02$\\
    Neural TS (Last Layer) + DeepFM & $1.72 \pm 0.13$ & $0.36 \pm 0.02$\\
    Ensemble + DeepFM & $0.021 \pm 0.001$ & $1.6 \pm 0.03$\\
    \midrule\midrule
    EpiNet + MLP & $0.013 \pm 0.001$ & $0.28 \pm 0.01$\\
    EpiNet + NCF & $0.015 \pm 0.001$ & $0.37 \pm 0.01$\\
    EpiNet + Wide \& Deep & $0.015 \pm 0.001$ & $0.38 \pm 0.01$\\
    EpiNet + DeepFM & $0.014 \pm 0.001$ & $0.38 \pm 0.01$\\
    \midrule
    ENR & $0.012 \pm 0.001$ & $0.38 \pm 0.01$\\
  \bottomrule
\end{tabular}
\vspace{-0.1in}
\end{table}

In this section, we will go through a set of experiments ranging from a toy experiment to large-scale real-world data experiments. The dataset statistics are shown in Table \ref{tab:stats}. Each of the dataset has millions of interactions. We will compare ENR and EpiNet \cite{osband2021epistemic} (first time adopted for neural contextual bandits) against $\epsilon$-Greedy, Ensemble Sampling with Prior Networks (Ensemble) \cite{lu2017ensemble}, Neural Thompson Sampling (Neural TS) \cite{zhang2021neural}, Neural UCB \cite{zhou2020neural}, and Neural LinUCB \cite{xu2021neural}, each with a similar sized neural network. In addition, to assess advantages attributable to the architecture of ENR, we will also evaluate it against a few well recognized RS neural network architecture including, Neural Collaborative Filtering (NCF) \cite{he2017neural}, Neural Recommendation with Attentive Multi-View Learning (NAML) \cite{wu2019naml}, Neural Recommendation with Multi-Head Self-Attention (NRMS) \cite{wu2019neural}, Wide and Deep \cite{cheng2016wide}, and DeepFM \cite{guo2017deepfm}, combined with exploration strategies above. To ensure fairness of evaluation, we ensure that all neural network architectures other than vanilla MLP has a similar parameter size. See Table \ref{tab:param}. The algorithms all adopt ADAM \cite{kingma2015adam} for model optimization.  

Note that for both Neural UCB and Neural Thompson Sampling, due to their scalability, we are only able to experiment them with their last layer versions, where we only compute gradients of the last layer of the neural networks. The full Neural UCB and Neural Thompson Sampling agents require inverting matrices of sizes $10^6 \times 10^6$ (7.45 TB of memory and $O(10^{18})$ complexity) and $10^7 \times 10^7$ (745 TB of memory and $O(10^{21})$ complexity) for MIND and KuaiRec experiments respectively, which are intractable in most commercial machines. See more details of computation costs in Table \ref{tab:mind_comp} and \ref{tab:kuai_comp} respectively for two experiments and all experiments are performed on an AWS p4d24xlarge machine with 8 A100 GPUs, each GPU with 40 GB of GPU memory. NeuralLinUCB requires 6x and 3.5x inference time compared to our method in two real-world experiments with a similar training time.  Simplified versions of NeuralTS and NeuralUCB using only gradients from the last layer of NNs require 10x and 100x inference time compared to our method in these experiments. Lastly, Ensemble requires a similar inference time, but requires 5x training time compared to our method with parallel computing optimization. NeuralLinUCB, NeuralUCB, NeuralTS all require orders of magnitude higher inference cost and Ensemble requires orders of magnitude higher training cost.

We will use average reward as performance metric in the following experiments (average click-through rate and average user rating in MIND and KuaiRec respectively). We choose to not use classic metrics like NDCG or precision because they are supervised-learning metrics. In online bandits, the agent’s optimal strategy could try to explore and retrieve information instead of optimizing these metrics.  

\begin{table*}[t!]
\small
  \caption{Average Regret for Toy Experiments}
  \vspace{-0.1in}
  \label{tab:toy_regret}
  \begin{tabular}{ccccccccccc}
    \toprule
    Algorithm & Exploit & $\epsilon$-Greedy & Neural UCB & Neural LinUCB & Neural TS & Ensemble & EpiNet & ENR\\
    \midrule 
    Avg Regret & $0.10 \pm 0.006$ & $0.11 \pm 0.01$ & $0.11 \pm 0.007$ & $0.07 \pm 0.009$ & $0.075 \pm 0.008$ & $0.08 \pm 0.004$ & $0.064 \pm 0.003$ & $\mathbf{0.053 \pm 0.0006}$\\
  \bottomrule
\end{tabular}
\vspace{-0.1in}
\end{table*}

\begin{table*}[t!]
\small
  \caption{User Interactions (Thousands) Needed for Best Algorithm Candidates (Selected among Different Neural Network Architectures) to Reach Standard Performance and Savings Compared to Non-EpiNet Baselines}
  \vspace{-0.1in}
  \label{tab:sample_complexity}
  \begin{tabular}{c|cccccc|cc|cc}
    \toprule
    Algorithm & Exploit & $\epsilon$-Greedy & Neural UCB & Neural LinUCB & Neural TS & Ensemble & EpiNet & ENR & Savings\\
    \midrule 
    MIND ($0.2$ Click-Through Rate) & $\infty$ & $\infty$ & $277$ & $268$ & $184$ & $142$ & $145$ & \textbf{62} & $56\%$\\
    KuaiRec ($2.3$ User Rating) & N/A & $4.5$ & N/A & $4$ & N/A & $2.4$ & $2.1$ & \textbf{1.7} & $29\%$\\
  \bottomrule
\end{tabular}
\vspace{-0.1in}
\end{table*}

 \begin{figure}
     \centering
     \includegraphics[width=0.3\textwidth]{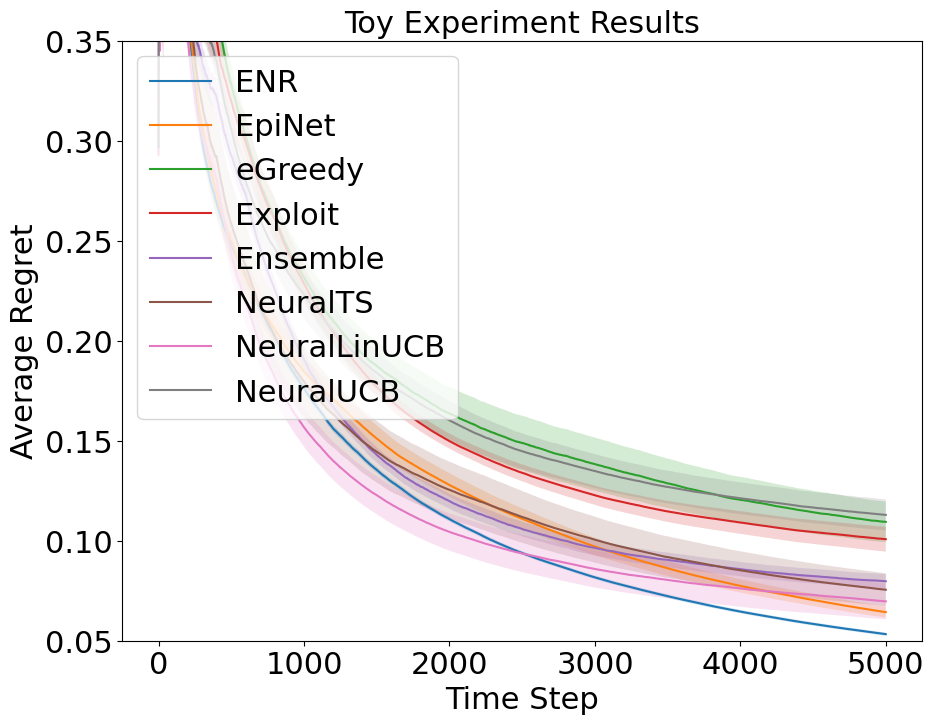}
     \caption{Toy Experiment Results}
     \label{fig:toy_1}
     \vspace{-0.1in}
 \end{figure}

\begin{table}[t!]
\small
  \caption{Average Click-Through Rate for MIND Experiments}
  \vspace{-0.1in}
  \label{tab:mind_reward}
  \begin{tabular}{ccccc}
    \toprule
    Algorithm & MIND Train & MIND Eval\\
    \midrule 
    Exploit + MLP & $0.170 \pm 0.0007$ & $0.168 \pm 0.0011$ \\
    $\epsilon$-Greedy + MLP  & $0.171 \pm 0.0007$ & $0.167 \pm 0.0011$\\
    Neural UCB (Last Layer) + MLP & $0.188 \pm 0.0016$ & $0.192 \pm 0.0010$\\
    Neural LinUCB + MLP & $0.189 \pm 0.0032$ & $0.192 \pm 0.00086$\\
    Neural TS (Last Layer) + MLP & $0.191 \pm 0.0023$ & $0.194 \pm 0.0015$\\
    Ensemble + MLP & $0.193 \pm 0.0022$ & $0.191 \pm 0.0009$\\

    \midrule
    Exploit + NCF & $0.181 \pm 0.0022$ & $0.179 \pm 0.0024$\\
    $\epsilon$-Greedy + NCF & $0.182 \pm 0.0023$ & $0.178 \pm 0.0025$\\
    Neural UCB (Last Layer) + NCF& $0.203 \pm 0.0029$ & $0.203 \pm 0.0023$\\
    Neural LinUCB + NCF & $0.203 \pm 0.0029$ & $0.203 \pm 0.0030$\\
    Neural TS (Last Layer) + NCF & $0.207 \pm 0.0031$ & $0.206 \pm 0.0028$\\
    Ensemble + NCF & $0.209 \pm 0.0040$ & $0.206 \pm 0.0031$\\
    
    \midrule
    Exploit + NAML & $0.180 \pm 0.0021$ & $0.178 \pm 0.0024$\\
    $\epsilon$-Greedy + NAML & $0.182 \pm 0.0022$ & $0.178 \pm 0.0025$\\
    Neural UCB (Last Layer) + NAML & $0.202 \pm 0.0028$ & $0.203 \pm 0.0024$\\
    Neural LinUCB + NAML & $0.203 \pm 0.0028$ & $0.203 \pm 0.0030$ \\
    Neural TS (Last Layer) + NAML & $0.206 \pm 0.003$ & $0.207 \pm 0.0028$\\
    Ensemble + NAML & $0.208 \pm 0.0039$ & $0.206 \pm 0.0031$\\

    \midrule
    Exploit + NRMS & $0.181 \pm 0.0022$ & $0.178 \pm 0.0024$\\
   $\epsilon$-Greedy + NRMS & $0.182 \pm 0.0023$ & $0.178 \pm 0.0025$\\
    Neural UCB (Last Layer) + NRMS & $0.203 \pm 0.0029$ & $0.203 \pm 0.0024$\\
    Neural LinUCB + NRMS & $0.204 \pm 0.0029$ & $0.203 \pm 0.0030$\\
    Neural TS (Last Layer) + NRMS & $0.206 \pm 0.0030$ & $0.207 \pm 0.0027$\\
    Ensemble + NRMS & $0.208 \pm 0.0039$ & $0.206 \pm 0.0031$\\

    \midrule\midrule
    EpiNet + MLP & $0.204 \pm 0.003$ & $0.199 \pm 0.0012$\\
    EpiNet + NCF & $0.219 \pm 0.004$ & $0.214 \pm 0.0028$\\
    EpiNet + NAML & $0.217 \pm 0.004$ & $0.214 \pm 0.0028$\\
    EpiNet + NRMS & $0.218 \pm 0.004$ & $0.214 \pm 0.0027$\\
    \midrule
    ENR & $\mathbf{0.228 \pm 0.004}$ & $\mathbf{0.220 \pm 0.0026}$\\
  \bottomrule
\end{tabular}
\vspace{-0.15in}
\end{table}

\begin{table}[t!]
\small
  \caption{Average Reward for KuaiRec Experiments}
  \vspace{-0.1in}
  \label{tab:kuai_reward}
  \begin{tabular}{ccccc}
    \toprule
    Algorithm & KuaiRec Train & KuaiRec Eval\\
    \midrule 
    $\epsilon$-Greedy + MLP & $1.41 \pm 0.087$ & $1.57 \pm 0.11$\\
    Neural LinUCB + MLP & $1.68 \pm 0.15$ & $1.59 \pm 0.13$\\
    Ensemble + MLP & $2.28 \pm 0.061$ & $2.03 \pm 0.09$\\
    \midrule
    $\epsilon$-Greedy + NCF & $2.42 \pm 0.0075$ & $2.31 \pm 0.019$\\
    Neural LinUCB + NCF & $2.47 \pm 0.022$ & $2.38 \pm 0.020$\\
    Ensemble + NCF & $2.55 \pm 0.003$ & $2.46 \pm 0.005$\\
    \midrule
    $\epsilon$-Greedy + Wide \& Deep & $2.38 \pm 0.007$ & $2.27 \pm 0.018$\\
    Neural LinUCB + Wide \& Deep & $2.42 \pm 0.021$ & $2.35 \pm 0.017$\\
    Ensemble + Wide \& Deep & $2.52 \pm 0.003$ & $2.43 \pm 0.005$\\
    \midrule
    $\epsilon$-Greedy + DeepFM & $2.38 \pm 0.0075$ & $2.32 \pm 0.022$\\
    Neural LinUCB + DeepFM & $2.45 \pm 0.022$ & $2.36 \pm 0.019$\\
    Ensemble + DeepFM & $2.53 \pm 0.003$ & $2.47 \pm 0.005$\\
    \midrule\midrule
    EpiNet + MLP & $2.43 \pm 0.01$ & $2.17 \pm 0.087$\\
    EpiNet + NCF & $2.58 \pm 0.0047$ & $2.43 \pm 0.016$\\
    EpiNet + Wide \& Deep & $2.56 \pm 0.005$ & $2.41 \pm 0.018$\\
    EpiNet + DeepFM & $2.57 \pm 0.0046$ & $2.46 \pm 0.021$\\
    \midrule
    ENR & $\mathbf{2.62 \pm 0.0005}$ & $\mathbf{2.63 \pm 0.066}$\\
  \bottomrule
\end{tabular}
\vspace{-0.2in}
\end{table}

\subsection{Toy Experiment}
For this experiment, we define a synthetic environment to evaluate the performance of ENR compared to various baselines mentioned above. According to Section \ref{sec:prob}, the synthetic environment is characterized as: Both action and context vectors are of dimension 100. $R_{t+1} \sim \text{Bernoulli}\left(\sigma(\theta^\top\text{concat}(\psi_{S_t}, \phi_{A_t}))\right)$. The size of $\mathcal{A}$ is 100. $\sigma$ is the Sigmoid function and both $\theta$ is unknown to all the agents as the learning target. All of $\theta, \phi_A, \psi_S$ are sampled from standard Gaussian. Since the toy environment is synthesized, we could compute the regret of the algorithms as
\begin{equation}
    \text{Regret}(T) = \frac{1}{T}\sum_{t=1}^T (R_{t, A^*_t} - R_{t, A_t}),
\end{equation}
where $R_{t, A^*_t}$ indicates the reward of the optimal action given the context $S_t$. In this experiment, we use a hidden layer of (200, 100) for MLP and the main neural network architecture for MLP and ENR. The action and context summarization layer for ENR is set without hidden layers and with activation of ReLU with Layer Normalization on top of that. The uncertainty estimation architecture of ENR is set with hidden layer of (50, 20). The EpiNet architecture is set to a single hidden layer of 50. All prior network scaling is set to 0.3. Dimension of epistemic indices is set to 5 for EpiNet and ENR. The results are averaged over 50 independent experiments. The experiment is run for 5000 steps.

See Figure \ref{fig:toy_1} and Table \ref{tab:toy_regret} for more details of the experiment results. We see that ENR marginally improves on top of EpiNet with a similarly sized base neural network of fully connected layers. Given that the setups of both toy experiments are relatively simple, it is expected that sophisticated neural representations would not make material difference.  

\begin{figure*}[ht!]
     \begin{subfigure}[t]{0.24\textwidth}
         \centering
         \includegraphics[width=\textwidth]{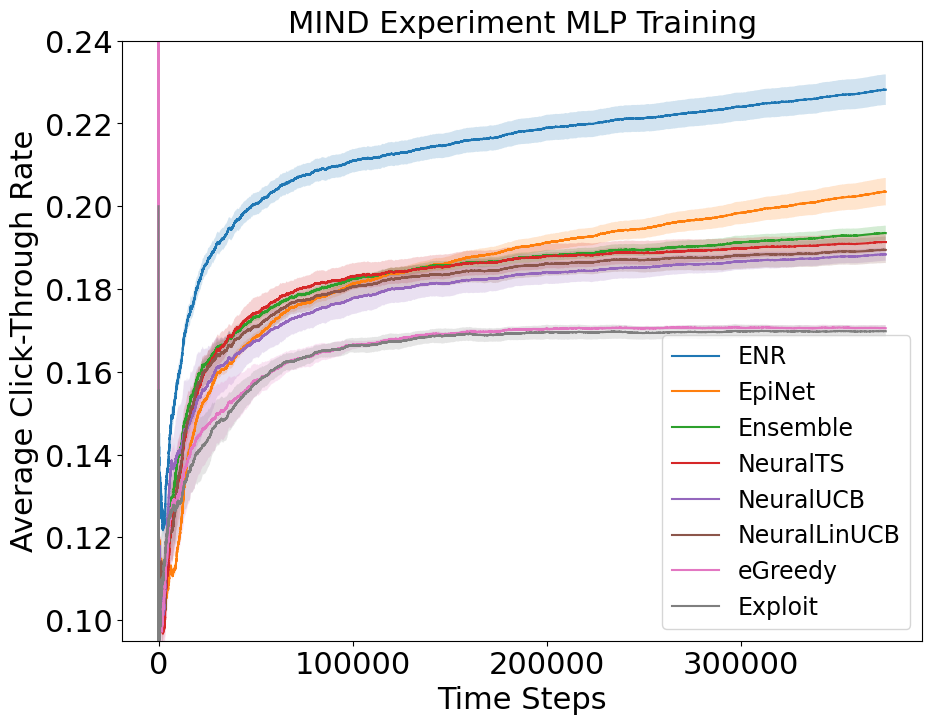}
         \caption{MIND Experiment MLP Training Results}
         \label{fig:mind_mlp}
         \vspace{-0.15in}
     \end{subfigure}
     \hfill
     \begin{subfigure}[t]{0.24\textwidth}
         \centering
         \includegraphics[width=\textwidth]{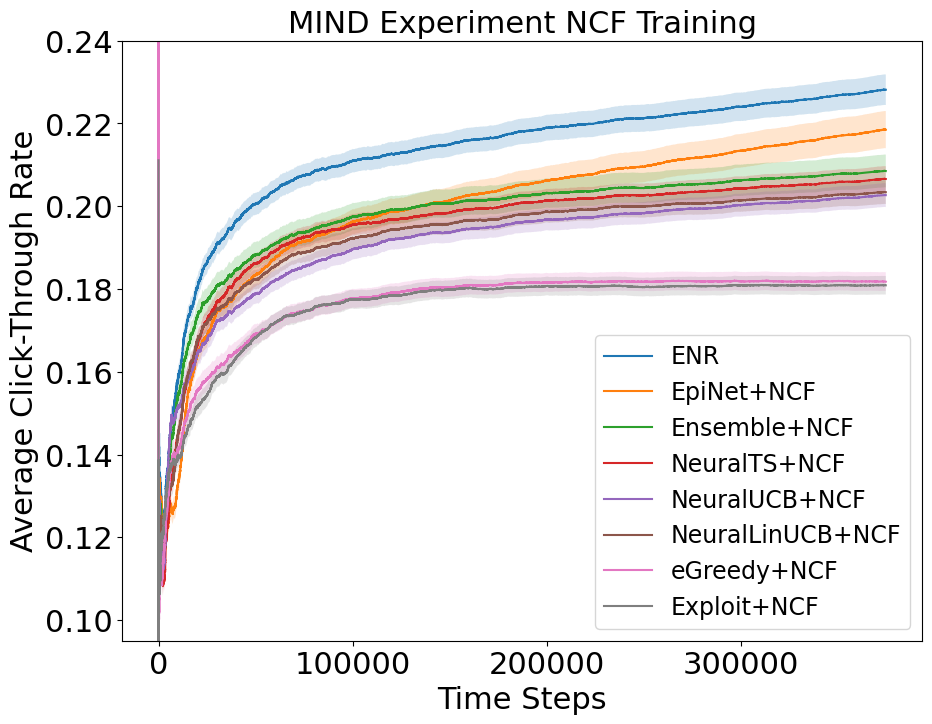}
         \caption{MIND Experiment NCF Training Results}
         \label{fig:mind_ncf}
         \vspace{-0.15in}
     \end{subfigure}
     \hfill
     \begin{subfigure}[t]{0.24\textwidth}
         \centering
         \includegraphics[width=\textwidth]{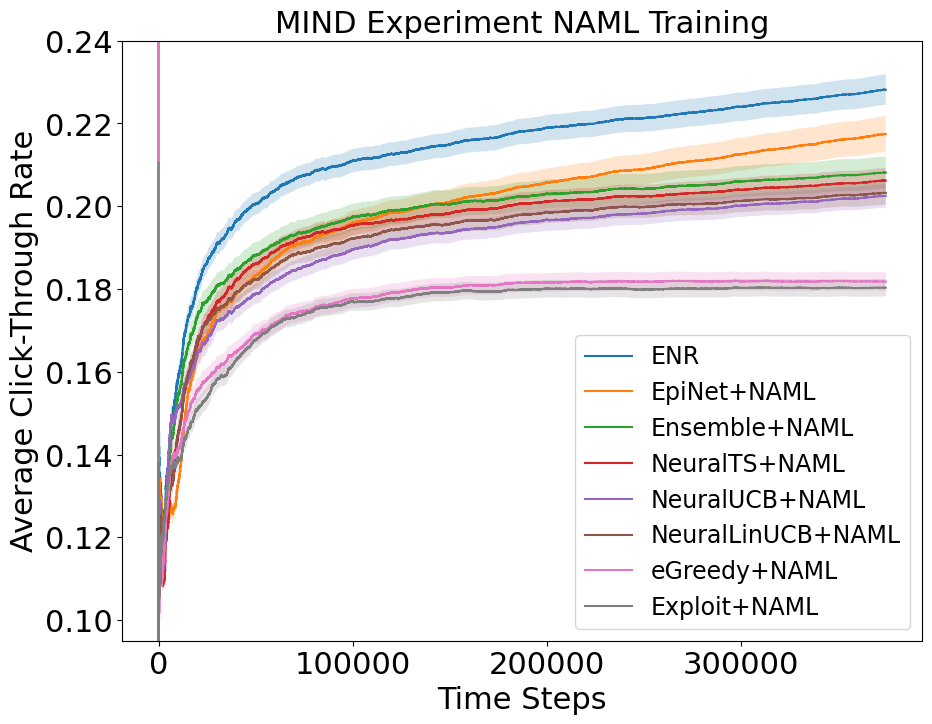}
         \caption{MIND Experiment NAML Training Results}
         \label{fig:mind_naml}
         \vspace{-0.15in}
     \end{subfigure}
     \hfill
     \begin{subfigure}[t]{0.24\textwidth}
         \centering
         \includegraphics[width=\textwidth]{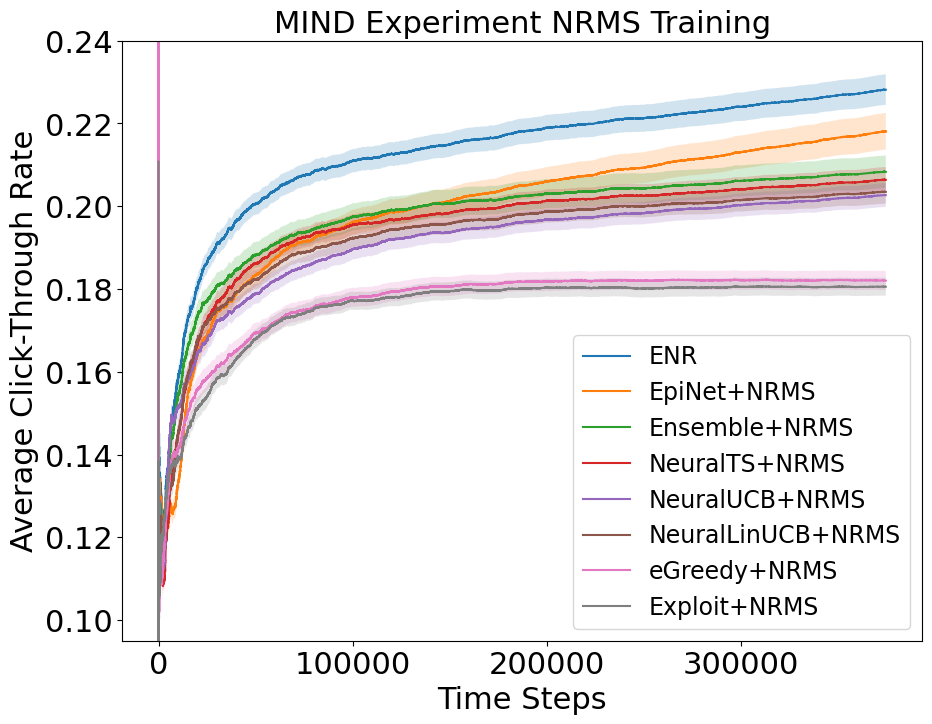}
         \caption{MIND Experiment NRMS Training Results}
         \label{fig:mind_nrms}
         \vspace{-0.15in}
     \end{subfigure}
     \caption{MIND Experiments Contextual Bandit Training Outcome}
     \label{fig:mind}
     \vspace{-0.1in}
\end{figure*}

\begin{figure*}[t!]
     \begin{subfigure}[t]{0.24\textwidth}
         \centering
         \includegraphics[width=\textwidth]{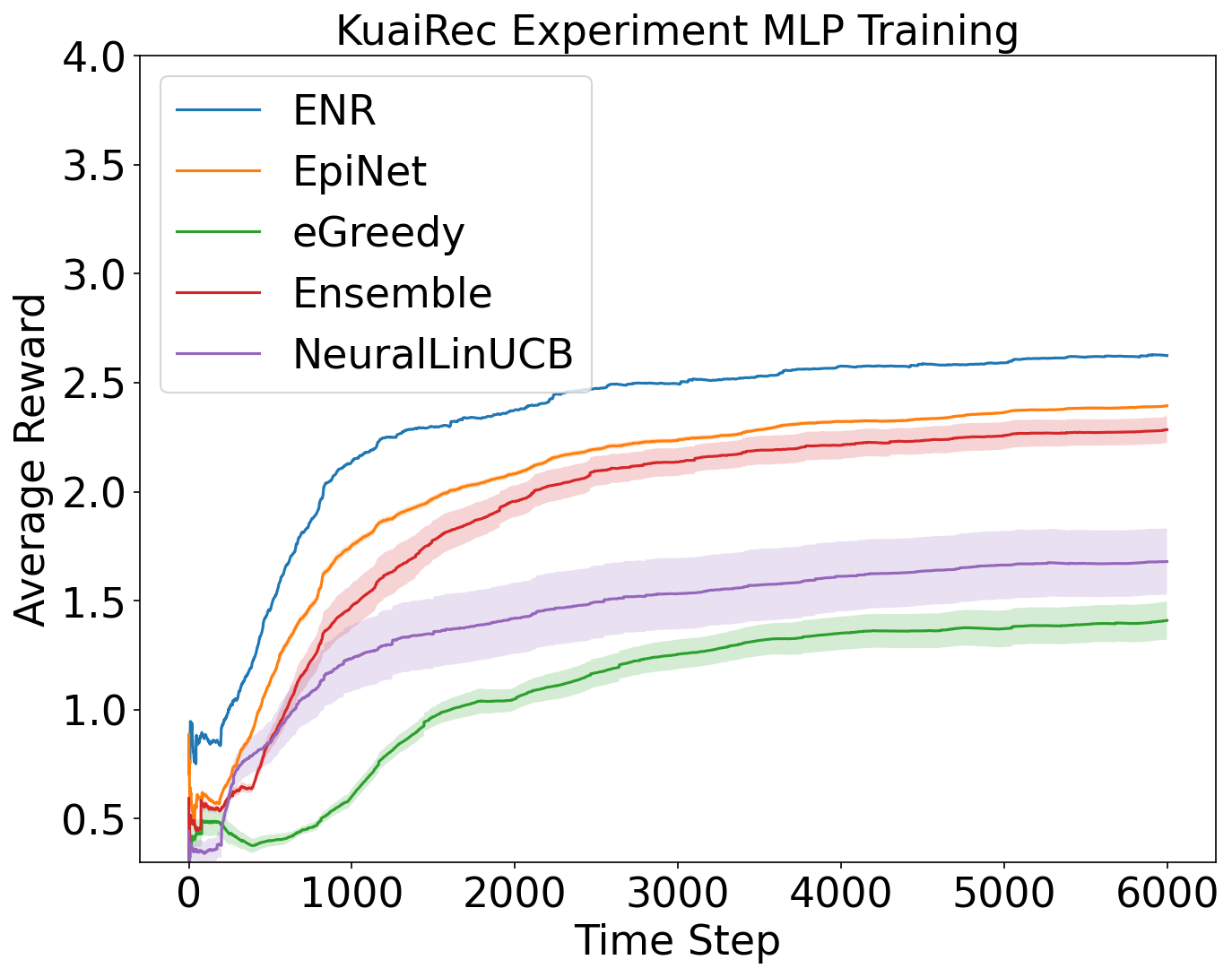}
         \caption{KuaiRec Experiment MLP Training Results}
         \label{fig:mind_mlp}
         \vspace{-0.15in}
     \end{subfigure}
     \hfill
     \begin{subfigure}[t]{0.24\textwidth}
         \centering
         \includegraphics[width=\textwidth]{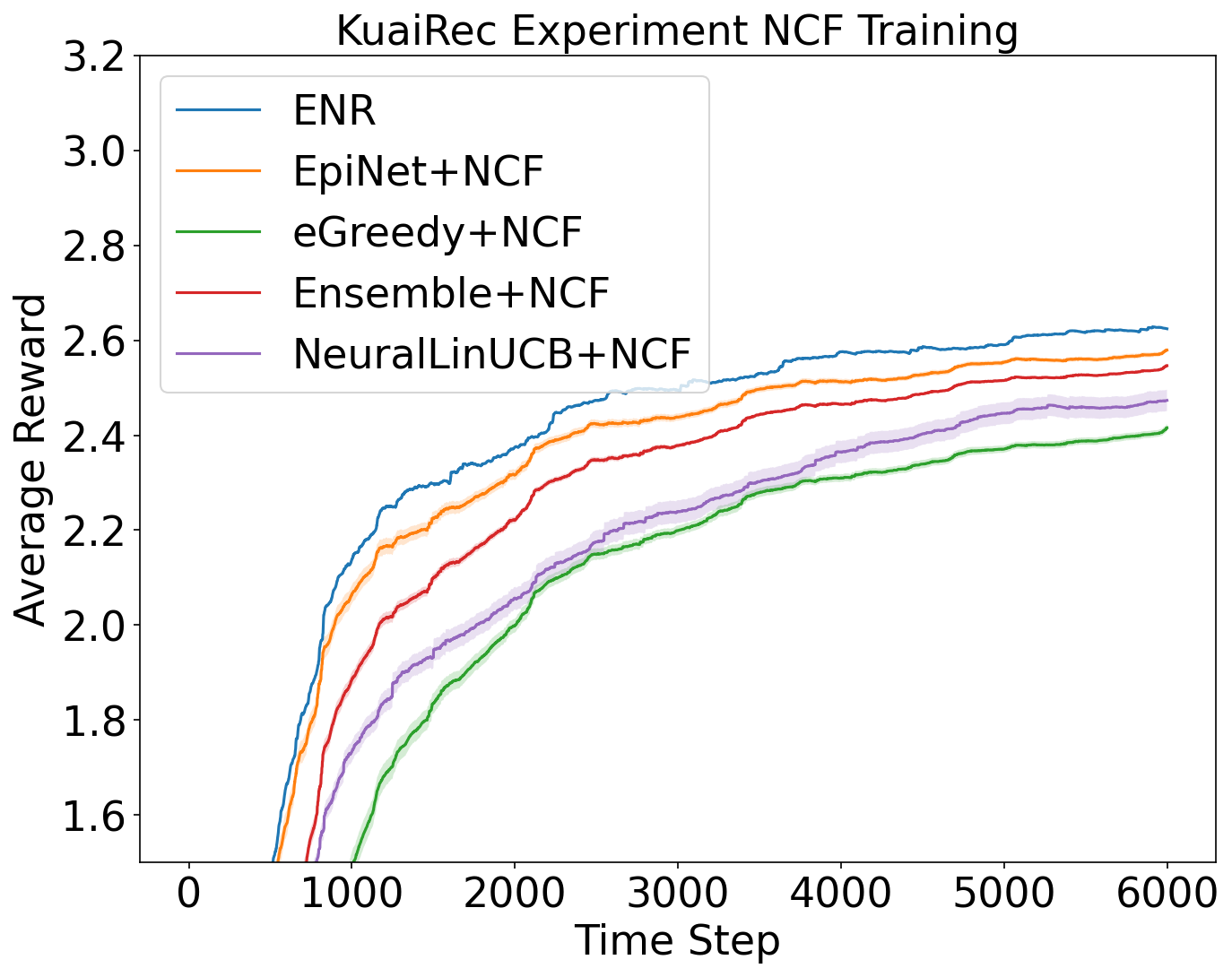}
         \caption{KuaiRec Experiment NCF Training Results}
         \label{fig:mind_ncf}
         \vspace{-0.15in}
     \end{subfigure}
     \hfill
     \begin{subfigure}[t]{0.24\textwidth}
         \centering
         \includegraphics[width=\textwidth]{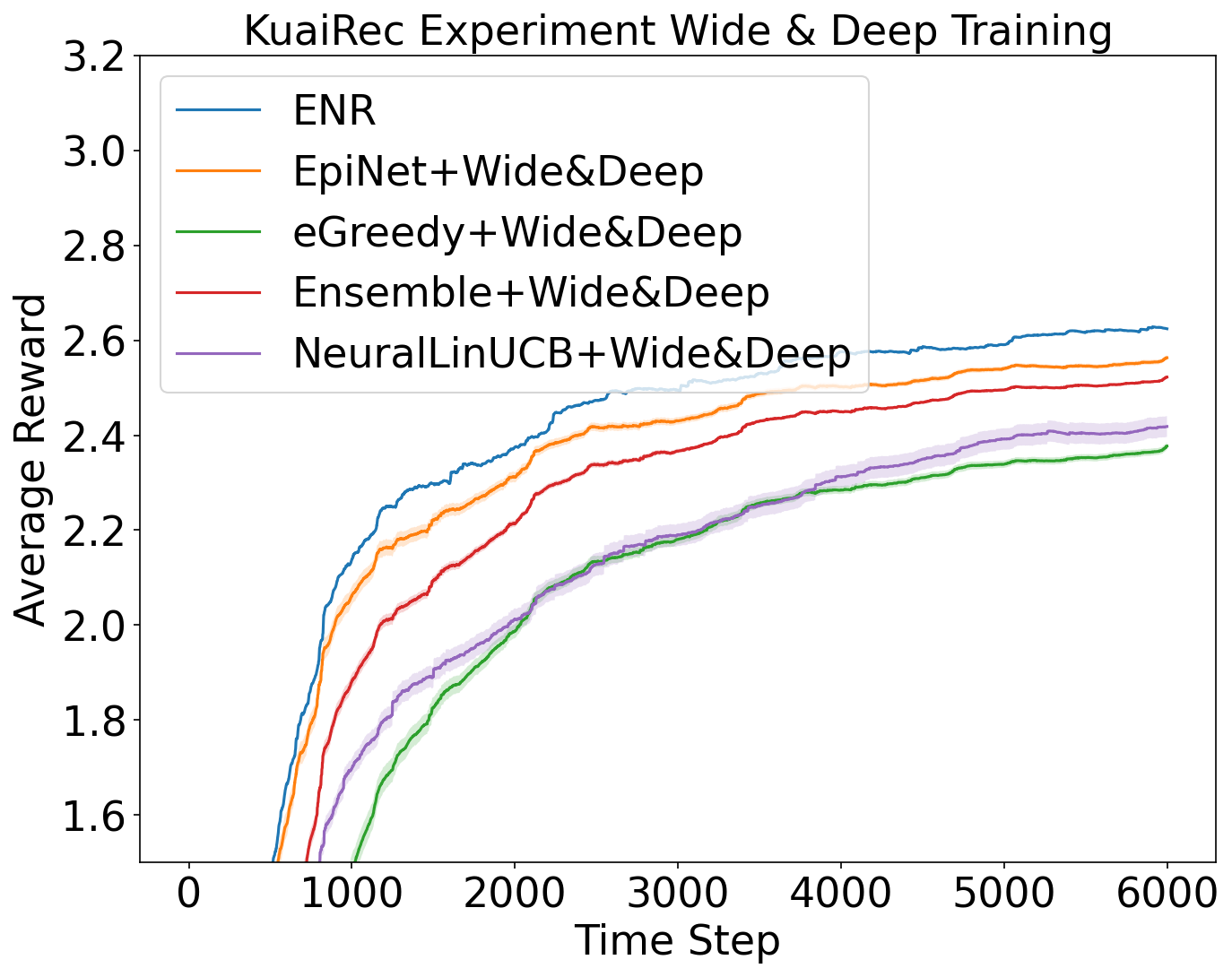}
         \caption{KuaiRec Experiment Wide \& Deep Training Results}
         \label{fig:mind_wd}
         \vspace{-0.15in}
     \end{subfigure}
     \hfill
     \begin{subfigure}[t]{0.24\textwidth}
         \centering
         \includegraphics[width=\textwidth]{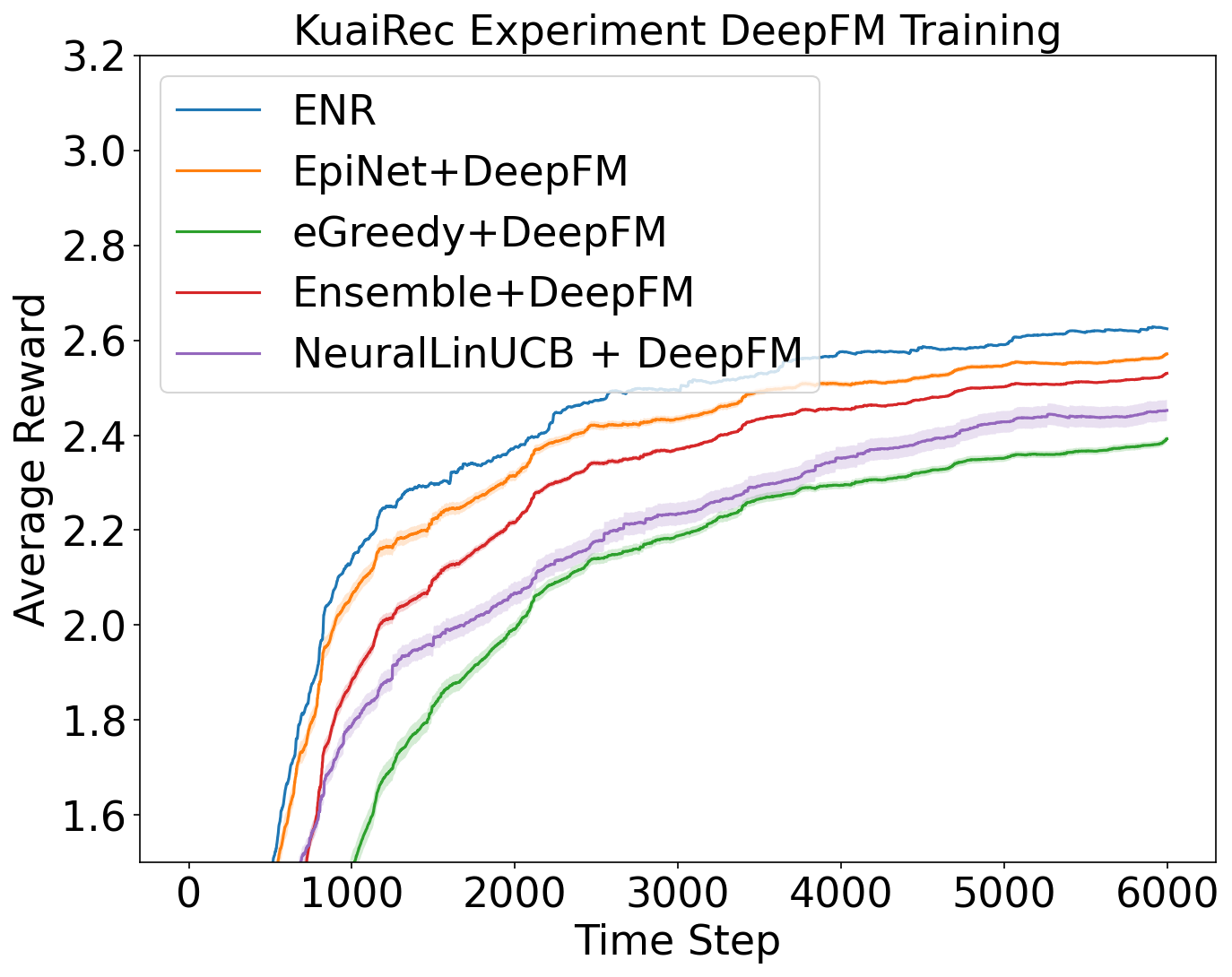}
         \caption{KuaiRec Experiment DeepFM Training Results}
         \label{fig:mind_deepfm}
         \vspace{-0.15in}
     \end{subfigure}
     \caption{KuaiRec Experiments Contextual Bandit Training Outcome}
     \label{fig:kuai}
     \vspace{-0.1in}
\end{figure*}

\subsection{MIND Dataset Experiment \cite{wu2020mind}}
MIND is a large-scale dataset generated by Microsoft news recommender system. It provides anonymized data logs that record each user's historical interesting articles and their impression as well as click logs with new recommendations. The dataset contains over millions of anonymized interaction logs. For statistics of the dataset, please refer to Table \ref{tab:stats}. One unique advantage of the MIND dataset for contextual bandit setup is its complete ground truth. At time step $t$, the MIND dataset logs $K$ distinct news recommendations sent to the user and records the news articles the user gave good feedback on, all of which happen at a single timestamp. Please see Table \ref{tab:mind_example} for an illustration of the data we use.  Hence when evaluating the outcome of any action chosen by an agent among the sent recommendations, we can directly use the groundtruth from the dataset and avoid counterfactual evaluation which leads to high variance. 

 In this experiment, we use all exploration baseline strategies as well as NCF \cite{he2017neural}, NAML \cite{wu2019naml} and NRMS \cite{wu2019neural} for neural network architecture baselines. It is worth noting that in this experiment, we adopt the Multi-Head Self-Attention architecture mentioned in NRMS \cite{wu2019neural} to extract representation for context (user) before feeding into the neural networks (except NAML where it has its own summarization module). We did not select Wide \& Deep and DeepFM because both neural architectures are heavily built on sparse feature optimization, which is not present in this experiment.

In the following, we use a hidden layer of (1000, 500, 100, 50) for MLP and the final neural network architecture for NCF and ENR. NCF's first MLP uses hidden layer of (1000, 500). For NAML and NRMS, EpiNet is added to the last layer replacing dot product with item-wise multiplication. The action and context summarization layer for ENR is set without hidden layers and with activation of ReLU and Layer Normalization on top of that. The uncertainty estimation architecture of ENR is set with hidden layer of (1000, 500). The EpiNet architecture for other neural networks are set with a single hidden layer of 50. All prior network scaling is set to 0.3. Dimension of epistemic indices is set to 5 for all EpiNets and ENR. The results are averaged over 10 independent experiments. Each training experiment is run for 375,000 time steps by sampling 375,000 users randomly from the dataset. In addition, we also sample 2,000 users that are never seen by the agent as evaluation set for out of sample evaluation.

See Figure \ref{fig:mind} as well as Table \ref{tab:mind_reward} for the experiment results. ENR is able to outperform all other baselines with a big margin in both online interaction and offline out-of-sample evaluation, outperforming $9\%$ and $6\%$ respectively in Click-Through Rate compared to the best baseline candidate (excluding EpiNet candidates, as that is considered part of the contribution of this work) respectively. Moreover, ENR is able to achieve a standard performance of $0.2$ click-through rate with $56\%$ less user interactions compared to the best non-EpiNet baseline. See Table \ref{tab:sample_complexity} for more details.

\begin{figure*}[t!]
     \begin{subfigure}[t]{0.3\textwidth}
         \centering
         \includegraphics[width=\textwidth]{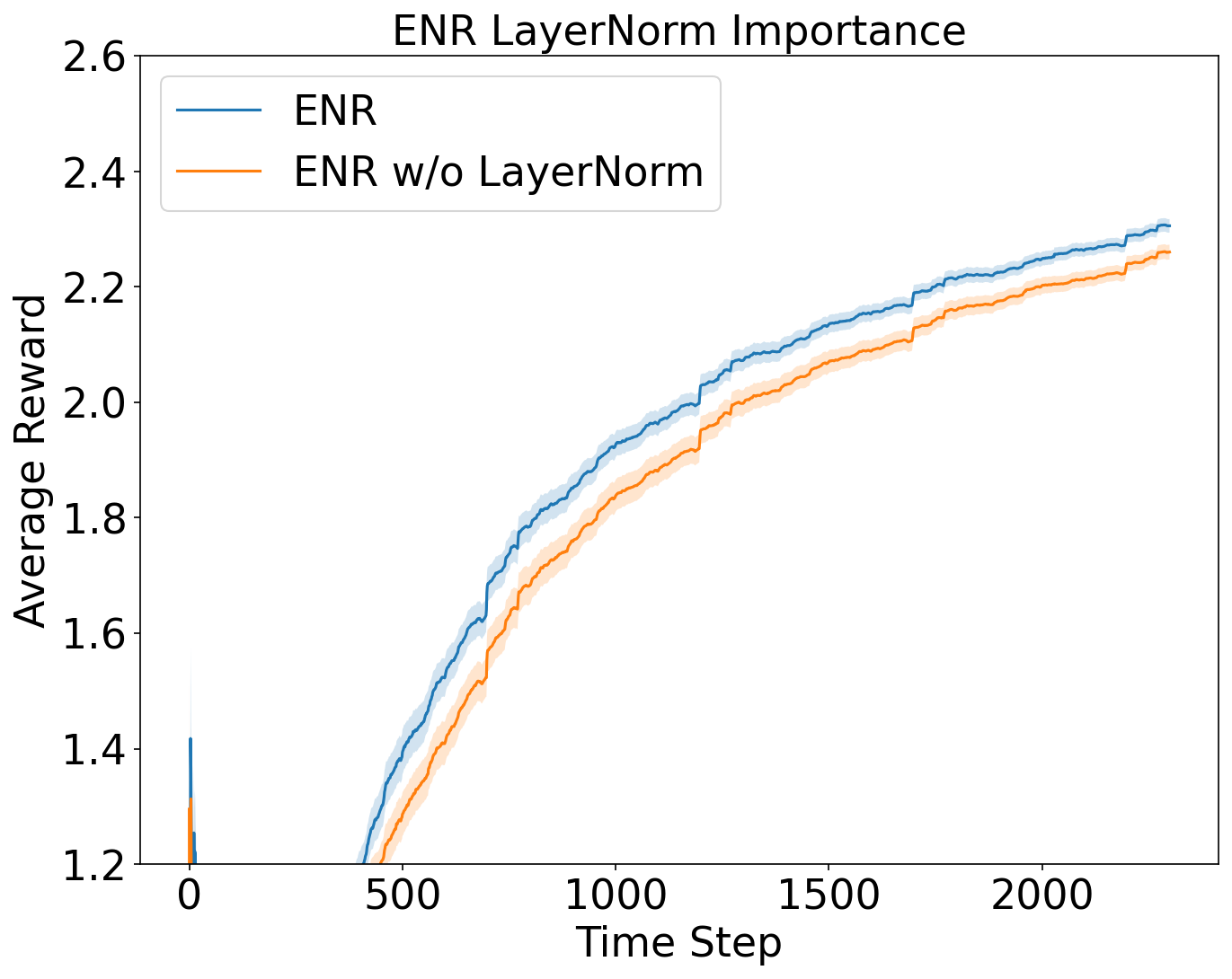}
         \caption{Comparison between with and without Layer Normalization Based on KuaiRec}
         \label{fig:layernorm}
         \vspace{-0.15in}
     \end{subfigure}
     \hfill
     \begin{subfigure}[t]{0.3\textwidth}
         \centering
         \includegraphics[width=\textwidth]{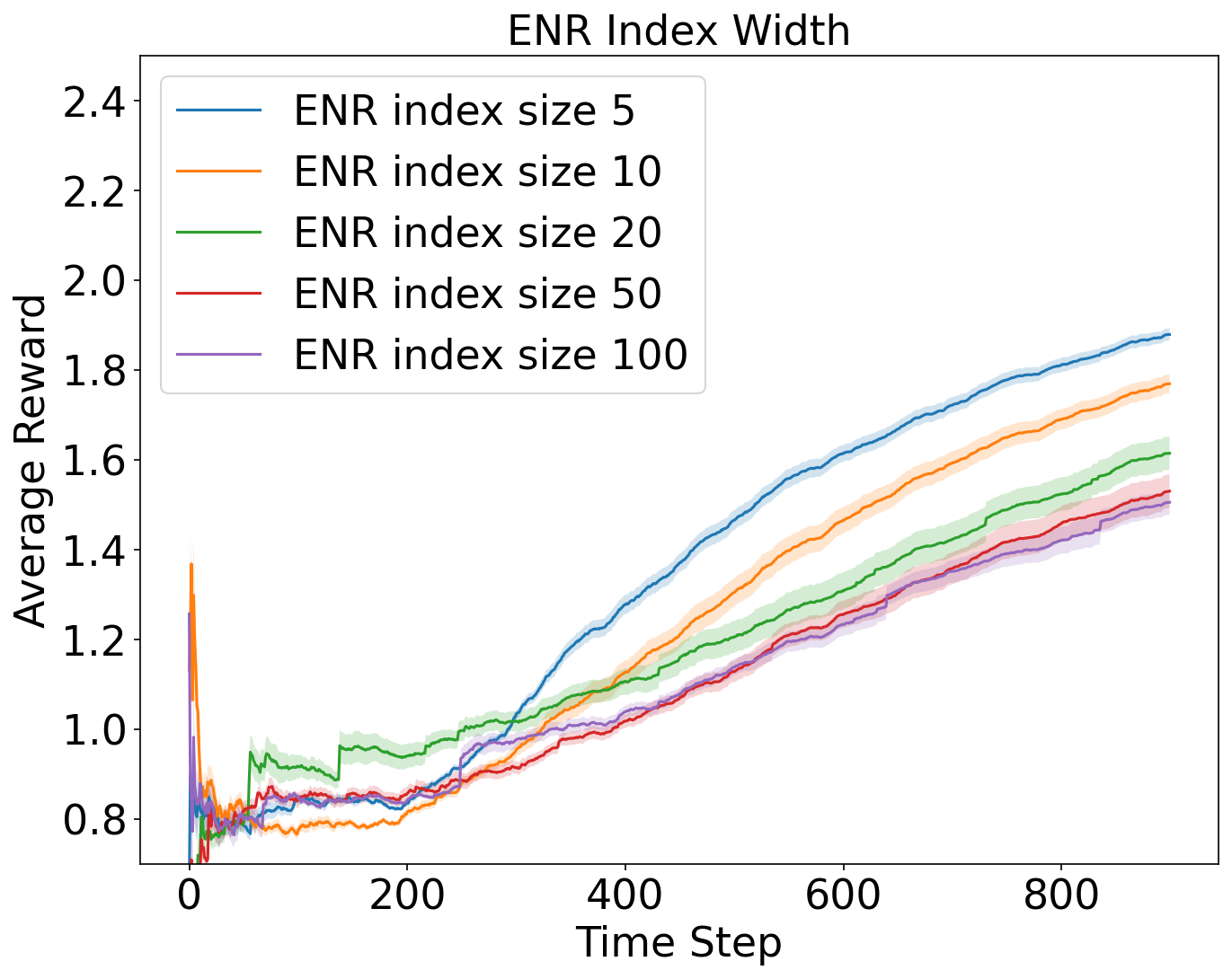}
         \caption{Selection of Index Width Based on KuaiRec}
         \label{fig:width}
         \vspace{-0.15in}
     \end{subfigure}
     \hfill
     \begin{subfigure}[t]{0.3\textwidth}
         \centering
         \includegraphics[width=\textwidth]{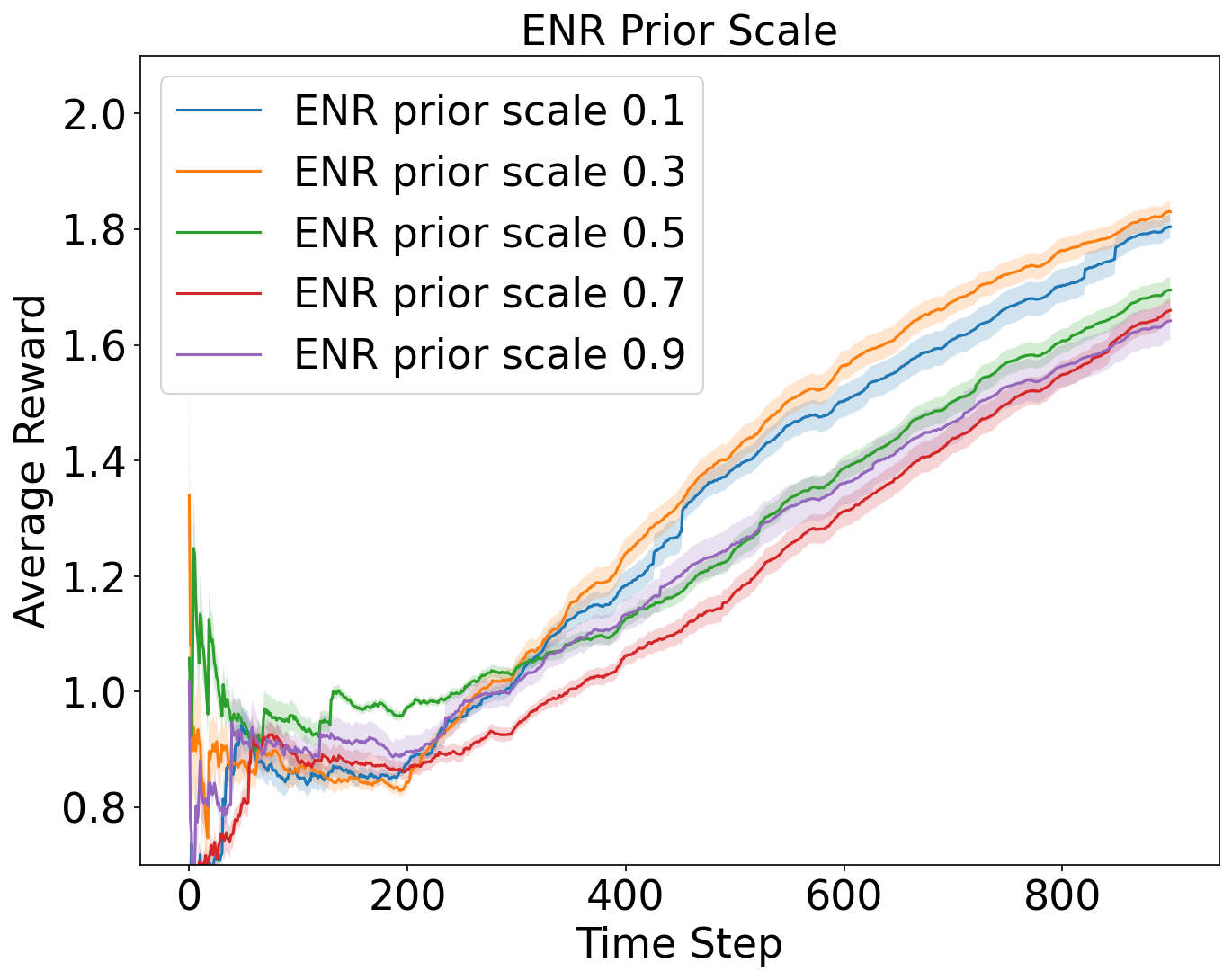}
         \caption{Selection of Prior Scale Based on KuaiRec}
         \label{fig:scale}
         \vspace{-0.15in}
     \end{subfigure}
     \caption{Ablation Studies}
     \label{fig:ablation}
     \vspace{-0.2in}
\end{figure*}

\subsection{KuaiRec Dataset Experiment \cite{gao2022kuairec}}
The last experiment we run is through the KuaiRec dataset. KuaiRec offers almost a full dense matrix in terms of the interactions between users and actions. The density is at 99.6\%. For more details of the dataset statistics, please refer to Table \ref{tab:stats}. With these characteristics of the dataset, we can evaluate all of the algorithms in real-world contextual bandit environment with huge available set of recommendations at every step without resorting to counterfactual evaluation. It is worth noting that, different from MIND, KuaiRec's user ratings on each recommendation are collected at slightly different times, and hence present a less rigorous grounding for avoiding counterfactual evaluation. Nevertheless, we believe it is valuable to still study the performance of ENR and EpiNet algorithms when facing a huge available set of actions at every timestep. We also note that such datasets, with feedback collected at different times, are commonly used in bandit learning literature to evaluate algorithms' performance, e.g. \cite{chris2018learning, hong2020latent, zhang2020conversational}.

In this experiment, we adopt all exploration strategies except Neural Thompson Sampling and Neural UCB due to their computation complexity. Given that both strategies requires a gradient pass on every action from the candidate list, these algorithms do not scale to a problem with 3,327 actions at every time step. We also choose NCF, Wide \& Deep and DeepFM as neural network architectures given that the features provided in the dataset are raw features with many categorical features. The dataset does not provide user historical articles and hence NAML and NRMS are not good fits.

In the following, we use a hidden layer of (4096, 1024, 512, 128) for MLP and the final neural network architecture for NCF and ENR. NCF's first MLP uses hidden layer of (1024, 512). Wide \& Deep and DeepFM adopts the same hidden layer structure for dense embeddings. The action and context summarization layer for ENR is set without hidden layers and with activation of ReLU with Layer Normalization on top of that. The uncertainty estimation architecture of ENR is set with hidden layer of (1024, 128). The EpiNet architecture for other neural networks are set with a single hidden layer of 128. All prior network scaling is set to 0.3. Dimension of epistemic indices is set to 5 for all EpiNets and ENR. The results are averaged over 10 random seeds. Each experiment is run for 6,000 time steps where each context is randomly sampled from the small matrix excluding an evaluation set of 100 contexts. One unique thing to note in this dataset is that user feedback for items in this dataset are rating scalars. Therefore, we use regression and mean squared error loss instead of cross entropy loss.

See Figure \ref{fig:kuai} as well as Table \ref{tab:kuai_reward} to see the detailed results of the experiment. From the results, we can see that ENR outperforms all other candidates in both in-sample online interactions as well as offline out-of-sample evaluation set by $4\%$ and $6\%$ respectively. Another observation we draw from the result is that with a complete action space provided to the agent at every timestep, all algorithms significantly improve in generalization capabilities from in-sample predictions to out-of-sample predictions. In Table \ref{tab:sample_complexity}, we also show that ENR requires $29\%$ less user interactions compared to the best non-EpiNet baseline to achieve a standard performance of $2.3$ average user rating. 

\vspace{-0.1in}
\subsection{Ablation Studies}
In this section, we cover some ablation studies regarding some of our hyperparameter choices as well as some design choices.
\vspace{-0.05in}
\subsubsection{The Importance of Layer Normalization}
In ENR, one of a key components for action and context summarization is Layer Normalization. As ENR is designed to take in any context and action inputs, it is common to observe a few features that are outsized. To ensure that context action interaction is not dominated by a few particular features, before the item-wise multiplication, we add a Layer Normalization operator to ensure more stability in the neural network as well as getting less impacted by outliers during training. See Figure \ref{fig:layernorm} for a performance comparison between ENR with and without Layer Normalization.
\vspace{-0.05in}
\subsubsection{Selection of Epistemic Index Width}
Epistemic index $z$ is the key to identify a posterior sample from an ENR, and we study the impact of index dimension in this study. See Figure \ref{fig:width}. As the width of the index grows, the performance of the algorithm relatively decays. From the figure, we can also see that index width of 5 outperforms others. Hence we choose 5 for all of our experiments.
\vspace{-0.05in}
\subsubsection{Selection of ENR Uncertainty Prior Network Scale} 
The prior network is a key means to regularize Epistemic Neural Networks by pulling the posterior distribution back from diverging too far away from the prior distribution to avoid distribution collapse. We study the prior network scale for ENR through a sweep. Note that prior scale is a scalar ranging $(0, 1)$. See Figure \ref{fig:scale}. We see that setting prior scale to 0.3 offers the best performance.

% \section{Discussion and Future Work}
\section{Conclusion}
In this paper, we designed a scalable and novel neural contextual bandit algorithm customized for recommender systems via a new epistemic neural network architecture and Thompson sampling. We formally define the recommender system problem as a contextual bandit problem and reviewed the current State-of-the-Art neural contextual bandit strategies. Our architecture design, Epistemic Neural Recommendation (ENR), presents much better scalability compared to other neural contextual bandit strategies. We show empirically through both synthetic experiments as well as two large-scale real-world experiments that ENR outperforms all other baselines. We hope that the results and the design in this paper inspires adoption of Epistemic Neural Recommendation as well as neural contextual bandit approaches in real-world systems.

\bibliographystyle{ACM-Reference-Format}
\bibliography{main}

\end{document}